# Super-resolution ranging using a sub-terahertz self-injection-locked frequency-modulated radar

Hossein Naghavi[1★], Zainulabideen Khalifa[2], Hamad Alotaibi[2,4], Farzad Khoeini[2], James Gruber[2], Morteza Tavakoli Taba[2], Aditya Varma Muppala[2], Saghar Adler[2], Ali Mostajeran[3], Mohammed Aseeri[4], Andreia Cathelin[5], Ehsan Afshari[2]

[1]Department of Electrical and Computer Engineering, University of Washington, Seattle, WA, USA. [2]Department of Electrical and Computer Engineering, University of Michigan, Ann Arbor, MI, USA. [3]Zadar Labs, San Jose, CA, USA. [4]King Abdulaziz City for Science and Technology, Riyadh, Saudi Arabia. [5]STMicroelectronics, Crolles, France.

**Sub-terahertz (sub-THz) and terahertz (THz) frequency-modulated continuous-wave (FMCW) radars have opened a plethora of scientific and industrial applications, especially in the imaging field. While strong candidates for sub-THz/THz FMCW radar imagers are implemented using photonic methods, there is a desire to achieve the full integration and portability that only electronics offers. However, integrated electronic sub-THz/THz FMCW radars have significantly lower bandwidth ($\leq$100 GHz) than photonic-based radars, restricting the radar range resolution to the millimeter size ($\geq$1.5 mm). In addition, the electronic FMCW radar's broad bandwidth comes with increased phase noise of the radar transmitter, consequently degrading the radar range accuracy. Here, we present a sub-THz fully-integrated autodyne frequency-modulated (AFM) radar engaged in a self-injection locking (SIL) mechanism that fundamentally overcomes the abovementioned challenges of FMCW radars. The AFM radar supports an enormously wide effective bandwidth up to the terahertz sweep range by forming an intermediate frequency comb**

**spectrum in a quadratic receiver, unlocking the path for super-resolution ranging. Also, SIL significantly enhances the radar's transmitter phase noise, allowing high-accuracy range measurements. We theoretically describe and experimentally demonstrate the SIL operation of the AFM radar. The proposed radar experimentally achieves sub-millimeter range resolution and range accuracy of < 0.002%, enabling imaging from covered printed letters with micrometer features.**

Sub-THz and THz FMCW systems have been attracting growing interest from researchers in the fields of photonics[1–5] and electronics[6–13] in recent years. Photonic FMCW radars generate frequency chirps in the sub-THz/THz band by photo-mixing of tunable- and fixed-wavelength lasers in a photoconductive antenna (PCA)[3] or a uni-traveling-carrier photodiode (UTC-PD)[1]. These photonic radars can provide a broadband frequency chirp up to the THz sweep range[3,14], which achieves micron range resolution. However, these FMCW systems use bulky lasers and separated transmitter (TX) and receiver (RX) parts, which make them expensive and limit their integrability into compact systems. On the other hand, electronic FMCW radars utilize a multiplier chain[7,10] or a harmonic voltage-controlled oscillator (VCO)[6,9,15] for generating a frequency chirp at sub-THz/THz bands. In the case of the multiplier chain topologies, a stabilized external source generates a highly linear and low-noise chirp signal at microwave frequencies, which is up-converted to the sub-THz/THz frequencies by the multiplier chain. When operating at the sub-THz/THz band, the multiplier topology consumes more power with less efficiency and needs a large chip area which, with the requirement of external sources, inhibit the full integration and portability of the system[16]. The harmonic VCO configuration employs a millimeter-wave oscillator with harmonics in sub-THz/THz frequencies. The frequency chirp is extracted from the harmonics, thus providing a bandwidth that is enhanced by the harmonic number. The major advantage of harmonic VCOs is that they can be fully integrated onto a single chip, making them very compact and cost-effective. However, the drawback is that the large bandwidth severely degrades the oscillator's phase noise[9,17], which leads to poor radar range accuracy[18].

Phase-locked loop (PLL)[15,19] and self-referenced frequency stabilization (SRFS)[20,21] approaches have been suggested to improve the VCO's phase noise and frequency stability by providing low-frequency feedback from the VCO output to its control voltage. A reference low-phase-noise oscillator and low-noise current source are required in the PLL and SRFS feedback loop, respectively. Nonetheless, the sub-THz/THz PLL circuits suffer from a large division ratio, limited locking range of frequency dividers, and significant noise folding[22] in the feedback loop. Also, the available SRFS methods proposed in the sub-THz/THz band show poor performance in VCO phase noise[21] and bandwidth[20,23], restricting the applicability of these techniques in broadband sub-THz/THz radars. In addition to the mentioned problems of electronic FMCW radars in the sub-THz/THz band, the broadest bandwidth reported in the literature is about 386 GHz[24], obtained with a bulky multiband structure, and this still cannot compete with its photonic counterparts. To overcome these limitations, we propose a fully-integrated silicon-germanium (SiGe) bipolar-complementary-metal-oxide-semiconductor (BiCMOS) 250 GHz AFM radar[6] engaged in a SIL mechanism that significantly enhances the harmonic VCO phase noise and enormously broadens the effective bandwidth of the radar by the formation of intermediate frequency (IF) comb spectrum in the radar's receiver.

SIL, or passive injection-locking, is analogous to active injection-locking, as explained by Adler's equation[25,26], and it is well-studied in lasers and, to some extent, in electronic oscillators. In this method, the oscillator (laser) becomes injection-locked to itself by taking a portion of the output signal and re-injecting it into the oscillator (laser) after propagating through a long delay line[27,28] or a high-quality factor (high-$Q$) resonator (e.g., Fabry-Perot[29], fiber[28,30], whispering gallery mode[31,32]). As opposed to active injection-locking[33,34], this approach to injection-locking does not need any external reference oscillators. Moreover, unlike PLL and SRFS, which operate based on low-frequency feedback that needs frequency division or down-conversion in the feedback loop, the self-injection locked (SIL[a]) oscillator works based on high-frequency feedback and does not need to lower the output frequency. Therefore, the settling time for frequency stabilization of the SIL oscillator is significantly faster than PLL and SRFS methods[35]. In this article, we illustrate that, in an AFM radar, part of the reflected signal from the target returns to the VCO through the TX/RX antenna and engages the VCO in a SIL mechanism. It is shown that strong SIL due to large target reflections can significantly enhances the VCO phase noise, consequently improving the radar range accuracy[18]. We also manifest the improved range accuracy of the SIL AFM radar by using it in a high-precision imaging setup with micrometer features.

Furthermore, one of the essential features of the SIL AFM radar is its capability to measure the range of targets through self-mixing interferometry[27] by interacting the internal voltage of the radar's VCO with the backscattered target signals. This way of measuring the target range differs from conventional FMCW radars, where mixing down to IF happens in a separate mixer. The moderate injection of the backscattered signal to the VCO drives it out of normal operation, entering a nonlinear dynamic regime where unexpected behaviors such as instability, bistability, and hysteresis appear[36]. By linearly sweeping the frequency of the SIL AFM radar, it moves through a series of unstable frequency regions and generates a train of sharp pulses in the time-domain IF output, which creates a frequency comb in the IF spectrum after de-chirping and Fourier transformation. Unlike conventional FMCW radars, which produce a single frequency line in the IF spectrum for a single target, the SIL AFM radar generates many equidistant frequency lines. We prove that higher-order lines ($\geq 2$) in the frequency comb spectrum results from fictitious FMCW radars with effective bandwidths equal to sweeping bandwidth times the line order number. As a consequence, for large enough frequency line orders, the effective bandwidth can reach THz sweep ranges and achieves super-resolution. We experimentally demonstrate that the SIL AFM radar can readily reach sub-millimeter range resolutions.

[a] For brevity, we use SIL as the abbreviation for both "self-injection locking" and "self-injection locked" phrases.

**Sub-THz AFM radar operation.** The AFM radar is a VCO that simultaneously conducts the functions of signal generation and mixing the transmitted and reflected signals; therefore can be considered a self-mixing interferometer[27]. There is no separate pathway for the RX signal in an AFM radar, as the transmission and reception occur simultaneously on the autodyne circuit[37]. Therefore, the radar employs a single antenna for both transmitting and receiving parts. Fig. 1a shows the deployed tabletop setup for the sub-THz AFM radar. The radar chip (Fig. 1a inset) is fabricated on a 55nm SiGe BiCMOS process, and its radiation and reception occur through a modified on-chip slot antenna[6]. The on-chip antenna mostly radiates through the substrate of the silicon chip using a hemispheric silicon lens to prevent the formation of unwanted surface waves inside the silicon substrate of the chip. The permittivity difference between the silicon region ($\varepsilon_r = 11.9$) and air ($\varepsilon_r = 1$) allows about 70% power transfer from silicon to air. A Teflon lens collimates the radiated power to form a sub-THz beam before impinging on a corner reflector, the radar target with a reflectivity close to one. The Teflon lens also focuses the returned power from the target onto the polished surface of the silicon lens, which again transfers 70% of the reflected power to the on-chip antenna. For simplicity, we neglect the multiple reflections between the target and the silicon lens, as verified by measurements. Overall, the radiated sub-THz wave experiences a round trip delay $\tau_d$ and loss $L$, as depicted in Fig. 1b.

**Principle of sub-THz VCO SIL (static analysis).** First, we present the general principles of the VCO SIL in the AFM radar to the long delay $\tau_d$ between radar and target (Fig. 1b) and define the basic terms (Fig. 2a-c). As illustrated in Fig. 1b, the proposed sub-THz AFM radar comprises a VCO and a quadratic receiver. The VCO is modeled by a lossy tank and a negative resistance ($-G_m$) to compensate for the tank loss. The generated frequency of the free-running VCO is determined by its tank resonance frequency ($\omega_{LC} = 1/\sqrt{L_1 C_1}$) and can be adjusted by varying the control voltage ($V_{tune}$) of varactor $C_1$. A coupling network couples the VCO to the free-space delay line by $g_1$ and $g_2$ coupling coefficients. Here, $v_P(t)$ is the pump voltage on the VCO tank, $v_R(t)$ is the radio-frequency (RF) signal at the on-chip antenna port, and $v_O(t)$ and $i_O(t)$ are the IF signals after down-conversion by the quadratic receiver. The voltage $v_R(t)$ comprises forward $v_R^+(t)$ and backward $v_R^-(t)$ waves, where $v_R^+(t)$ is a copy of the pump signal $v_P(t)$ and $v_R^-(t)$ is a delayed version of pump signal $v_P(t-\tau_d)$. Here, the coupling network injects the delayed version of the pump signal $v_P(t-\tau_d)$ into the VCO and alters the VCO behavior due to SIL. In that case, $\omega$ is the effective VCO frequency that differs from $\omega_{LC}$, due to the sub-THz feedback from the long delay $\tau_d$, which impacts the VCO dynamics. In Fig. 2a-c, the SIL modeling is achieved from the static analysis of the equations of motions, presented in Supplementary.

Fig. 2a shows the stationary tuning curve (STC) for the dependency of free-running VCO frequency ($\omega_{LC}$), equivalently the control voltage ($V_{tune}$) of varactor $C_1$, versus the effective VCO frequency ($\omega$)[29–32]. In the absence of a target, which means no SIL, the tuning curve tracks the line $\omega_{LC} = \omega$ (Fig. 2a) as $V_{tune}$ varies. However, with the presence of a target, the VCO frequency becomes self-injection locked due to the feedback from the target backscatter. Strong enough feedback significantly deforms the STC (Fig. 2a, b) and the output power spectrum of the pump signal (Fig. 2c). In that case, for some ranges of the free-running frequency $\omega_{LC}$, there are three-solution regions for the effective frequency $\omega$, where always the middle branch is unstable[31], and the two other stable branches create a bistable area (Fig. 2a). The bistability in the tuning curve demonstrates hysteresis behavior in the up-chirping and down-chirping directions (Fig. 2b). Every time the free-running VCO frequency reaches a turning point in the STC, the effective frequency hops between stable states. Moreover, Fig. 2b illustrates the VCO locking range ($\Delta\omega_{lock}$), where its slope ($d\omega_{LC}/d\omega$) defines the stabilization factor for the SIL model[29,31]. The locking range width ($\Delta\omega_{lock}$) depends on the magnitude of the reflection coefficient $\Gamma_{in}(\omega)$, the coupling coefficients $g_1$ and $g_2$, and the antenna port impedance $Z_0$ (Fig. 1b).

As shown in Fig. 2a, inside the locking range, the fluctuation of the effective VCO frequency $\omega$ is several times smaller than the variation of the free-running VCO frequency due to the large stabilization factor ($d\omega_{LC}/d\omega$). Measurement results in Fig. 2d-f prove this observation. Fig. 2d shows the power spectrum of the sub-THz signal after down-conversion, which manifests a narrower linewidth and more substantial power for the SIL VCO compared to free-running VCO. Also, sub-THz feedback from the long delay $\tau_d$ suppresses the short-term statistical phase errors in the system, like the VCO phase noise[28]. As shown in Fig. 2e, the phase noise is reduced by more than 20 dB (at 1 MHz offset frequency) over a wide frequency range compared to the free-running VCO. Moreover, Fig. 2f compares the long-term frequency stability of free-running and SIL VCOs by plotting the Allen deviation and shows about 7 dB improvement. The instantaneous frequency measurement of Fig. 2f is reported in the Supplementary for a 1000-second measurement.

**Pulse train in the SIL AFM radar's quadratic receiver (dynamic analysis).** Before studying the pulse train formation at the IF port of the quadratic receiver (Fig. 1b), we review the dynamic behavior of SIL VCO when free-running frequency $\omega_{LC}$ moves through the unstable areas of the STC. Fig. 3a shows the measured up-chirping spectrogram of the SIL VCO. Here, the $V_{tune}$ of varactor $C_1$ is continuously varied to increase the free-running frequency of the radar's VCO over time; therefore, the time axis in Fig. 3a is proportionate to the free-running frequency similar to Fig. 2a,b. It demonstrates the instantaneous jumps of the VCO effective frequency $\omega$ between adjacent stable states of the STC (known as frequency hopping[38]). When the SIL VCO frequency reaches a turning point of the up-chirp locked state in the STC (Fig. 2b and Fig. 3a), it jumps through the unstable region and settles in the adjacent stable state. Fig. 3b illustrates the time-domain version of the measured radiated sub-THz signal of the SIL VCO and the instantaneous frequency increase during the hopping between two stable states. The dynamic analysis of the SIL VCO has been carefully conducted in the Supplementary with similar results matching measurements. Moreover, we observe the same behavior during the down-chirping of the SIL VCO, as shown in Fig. 3c. For the down-chirping, the $V_{tune}$ of the varactor $C_1$ is continuously changed to decrease the free-running frequency of the radar's VCO, and the time axis is in the opposite direction of the free-running frequency in Fig. 2a,b. As illustrated, the measured effective frequency moves along the other side of the STC (Fig. 2b), and it instantaneously decreases during the hopping between two stable states (Fig. 3d).

In the simplified circuit model of Fig. 1b, the nonlinear impedance $Z_N(v)$ carries out the function of down-conversion in the AFM radar's receiver. The impedance $Z_N(v)$ in the receiver encloses the dominant quadratic nonlinear terms of the VCO's transconductance $-G_m$ and varactor $C_1$. Therefore, the AFM radar's receiver is not a separate block from VCO; for simplicity, it is illustrated as an independent block. Here, the voltage across the nonlinear impedance $Z_N(v)$ has a high-frequency term $v_P(t)$ and a low-frequency term $v_O(t)$. The quadratic nonlinearity of $Z_N(v)$ mixes $v_P(t)$ and $v_O(t)$ in such a way that an instantaneous frequency change in $v_P(t)$ generates a copy in low frequencies that shows itself as a kink in the IF signals $v_O(t)$ and $i_O(t)$[39]. Based on the dynamic analysis of the SIL, developed in the Supplementary, the resistive and capacitive parts of $Z_N(v)$ generate a step function and an impulse in the IF current $i_O(t)$, respectively. However, it was observed from the measurement results that the dominant nonlinearity comes from the resistive part of $Z_N(v)$. Hence, each frequency hop produces a step jump in $i_O(t)$. As shown in Fig. 1b, $i_O$(t) is extracted from the biasing line using a bias-T, and after trans-impedance amplification, it passes through a differentiator which converts the step jump in $i_O(t)$ to a sharp pulse with a uniform broad bandwidth. Fig. 3b,d illustrate the processed IF pulses for up-chirping and down-chirping, respectively. From the measurements, the generated IF pulses during up-chirping are stronger than those during down-chirping. We should clarify that throughout several measurements, the IF pulses generated during down-chirping are not as stable as the IF pulses generated during up-chirping, and their pulse shapes vary depending on the target distance. Hence, moving forward, we will focus only on the up-chirping operation of the SIL AFM radar.

By linearly scanning the frequency of SIL VCO, the radiated frequency jumps between stable states of the STC, and for each frequency hop, the radar receiver delivers a sharp pulse at the IF port. Fig. 4a illustrates the IF output of the sub-THz SIL AFM radar with 67 GHz bandwidth for a target range of 30 cm and 10 kHz chirp rate. A broadband chirp linearization technique is applied to form a time-domain pulse train at the IF signal[40]. Fig. 4b shows the fast Fourier transform (FFT) of the time-domain pulse train, revealing a frequency comb spectrum with discrete and equidistant frequency lines. Here, the amplitude of the lines gradually drops due to the limited bandwidth of the IF circuitry (TIA and VGA in Fig. 1a) and the deficiencies in the pulse recovery process. Moreover, the density of frequency lines can be controlled by adjusting the radar chirp rate; the slower the chirp rate, the higher the line density. More discussions about the pulse recovery process and the benefits of adjusting the radar chirp rate are explained in the Supplementary.

**Super-resolution ranging**. As depicted in Fig. 4c, in an FMCW radar with a single target, the IF output spectrum has one frequency line, and its frequency is linearly proportional to the radar scanning bandwidth; $f_{IF} = (2Rf_m/c)B$, where $B$ is the VCO bandwidth of FMCW radar, $f_m$ is the chirp rate ($T_m = 1/f_m$ is the chirp period), $c$ is the speed of light in air, and $R$ is the target range. However, for the proposed SIL AFM radar with a single target, the frequency comb spectrum has many frequency lines depending on the target distance, pulse bandwidth, radar chirp rate, and bandwidth of the IF circuitry. Each frequency line in Fig. 4c is a harmonic of the first frequency line $f_{IF,N} = Nf_{IF,1}$, and its frequency is linearly proportional to the radar scanning bandwidth ($B$) times the frequency line order ($N$); $f_{IF,N} = (2Rf_m/c)NB$. In this formula, $c$, $f_m$, and $R$ are the same for all frequency lines, and the only difference between the frequency lines is their *effective bandwidth* that is given by $NB$. As depicted in Fig. 4c, the $N^{th}$ frequency line in the IF spectrum of the SIL AFM radar can be interpreted as a conventional FMCW radar with an effective bandwidth that is $N$-times the scanning VCO bandwidth $B$. Therefore, we can model the SIL AFM radar with a synchronous ensemble of FMCW radars, where their bandwidths are $NB$.

Range resolution and accuracy are two fundamental nomenclatures in FMCW radars. The *range resolution* of an FMCW radar is the minimum distance that two targets can be separated along the radar's line of sight before they are indistinguishable. This feature is critical for concealed object detection, which requires separation between individual surfaces of imaging scenes[41]. The range resolution of a conventional FMCW radar depends only on the radar's bandwidth using $\Delta R_{\mathrm{min}} = \alpha c/(2B)$[41,42], where $\Delta R_{\mathrm{min}}$ is the range resolution, and $\alpha$ is the window factor of the IF signal due to amplitude modulation ($\alpha \geq 1$). However, as explained earlier, the SIL AFM radar works with higher-order frequency lines, and the effective bandwidth for the $N^{\mathrm{th}}$ frequency line in the IF spectrum is $N$-times larger than the first line bandwidth ($B$), readily reaching THz sweep ranges. To prove this quality of SIL AFM radar, a Fourier transform spectroscopy (FTS) method[43] has been applied to measure the radiated frequency of SIL AFM radar at each time, as displayed in Extended Data Fig. 3. This technique of achieving the FTS spectrogram has been discussed in the Supplementary. The $N^{\mathrm{th}}$ frequency line in the IF spectrum of Fig. 4c is the outcome of FMCW($N$) radar in the FTS spectrogram with effective bandwidth $NB$. We should emphasize that only FMCW(1) is the only physically radiating radar, and the rest of FMCW($N \geq 2$) radars are fictitious, created by the SIL nonlinearity. The measured radiated frequency of SIL AFM radar using a spectrum analyzer verifies this observation. Therefore, we can express the range resolution of the $N^{\mathrm{th}}$ frequency line of the SIL AFM radar by $\Delta R_{\mathrm{min}} = \alpha c/(2NB)$, where an integer $N$ is added to the denominator of the range resolution formula for a conventional FMCW radar. This ability of SIL AFM radar to look at higher-order frequency lines allows it to resolve objects much closer than the limit imposed by FMCW radars. Fig. 4d shows the first, ninth, and eleventh lines in the frequency comb spectrum when the SIL AFM radar looks at two objects $\Delta R = 570$ μm apart. The range resolution of the first reflection line ($N = 1$) is equal to the resolution of a conventional FMCW radar with 67 GHz bandwidth ($\Delta R_{\mathrm{min}} = 3$ mm for $\alpha = 1.33$), which cannot distinguish the two objects. However, the $9^{\mathrm{th}}$ and $11^{\mathrm{th}}$ frequency lines with respective effective bandwidths of 603 GHz (9×67 GHz) and 737 GHz (11×67 GHz) can readily resolve two objects since these frequency lines have range resolutions of $\Delta R_{min} = 330$ μm and $\Delta R_{min} = 271$ μm, respectively, which are smaller than the separation of the targets $\Delta R = 570$ μm. For comparison, each of the frequency line spectrum sub-figures contains two other measurements, each corresponding to the case where the radar looks at only one of the two targets.

**High-accuracy ranging and high-precision imaging.** On the other side, the *range accuracy* of radar is defined as how accurately the distance to a single target can be determined[41]. FMCW radars determine the target range by measuring the frequency or phase of the IF signal[6,37]. The radar's VCO phase noise affects the accuracy of frequency and phase measurements of the IF signal, directly introducing inaccuracies in determining the target range[18]. In traditional FMCW signal processing, the poor phase noise performance of the VCO is eliminated, to some degree, by signal averaging and correlation between the transmitted and received signals[41]. However, this is insufficient for high precision range measurements on the order of single-digit microns, which needs very low phase noise VCOs. As shown in Fig. 2e, the SIL technique significantly improves the AFM radar phase noise, enhancing the radar's range accuracy to 3.4 μm for a target at 185.5 mm (Fig. 5a). Moreover, the SIL technique enhances the frequency stability of the radar (Fig. 2f) which, in combination with the improved phase noise, enables the SIL AFM radar to be utilized in high-precision imaging. Fig. 5b illustrates the imaging setup from a printed letter "M" with a 12 μm ink thickness covered by *n* layers of paper. The metallic plate behind the papers is an intensely reflective target that puts the AFM radar in a deep SIL regime. By scanning the image scene using a 2*D* motorized stage, we can precisely measure the effective range variations due to changes in the refractive index of the areas with and without ink. Fig. 5c shows several images of the printed letter "M" when covered with several layers of paper. Due to the moderate transparency of the papers at sub-THz frequencies, even after 20 pages, the printed letter "M" is still readable[44]. The image SNR degrades with increasing paper layers due to the fluctuations in paper thickness and refractive index at different image locations.

Finally, Extended Data Fig. 4a tabulates the range accuracy of the radar for several ranges and demonstrates a range accuracy better than 0.002% for the SIL AFM radar when averaging 100 IF signals. Also, Extended Data Fig. 4b,c shows additional reconstructed images from printed "ABC" letters with different ink thicknesses. The image quality drastically improves with thicker ink.

**Conclusions.** We have demonstrated a fully-integrated sub-THz AFM radar engaged in a SIL regime to enormously enhance the performance of the radar compared to conventional FMCW radar systems. The static and dynamic analyses of the SIL process are presented. The SIL mechanism significantly reduces the radar's VCO phase noise, consequently improving the radar range accuracy to <0.002%. Moreover, sharp pulse generations during frequency hoppings generate a broad frequency comb in the IF spectrum, which results in super-resolution by working with higher-order frequency lines. We theoretically and experimentally proved that the range resolution limit of $\Delta R_{\min} = \alpha c/(2B)$ is not a constraint for the SIL AFM radar as it can obtain any desired resolution of $\Delta R_{\min} = \alpha c/(2NB)$ by appropriate selection of frequency line order $N$. However, this achievement comes with two limitations. First, the restricted frequency separation between the IF comb lines limits the range window of the SIL AFM radar. Therefore, for complex image scenes with many scatterers, if targets fall outside the range window, they will interfere with adjacent frequency lines. To tackle this issue, increasing the chirp rate expands frequency comb spacing and allows a larger range window for each frequency line.

Second, operating at large values of line order $N$ requires advanced chirp linearization methods at hardware and software levels to get the super-resolution privilege of SIL AFM radars. In summary, the proposed SIL AFM radar is an excellent fully-integrated candidate for super-resolution and high-precision imaging applications with micron capabilities.

## Methods

**Details of laboratory measurement setup in Fig. 1a.** In the RF measurements, we utilized a Rohde&Schawrz FSW67 spectrum analyzer (SA) with FS-Z220 and FS-Z325 even harmonic mixers (EHM) for RF power spectrum measurements. A Keysight MSOS 804A high-definition oscilloscope (RF Osc) accompanied by two cascaded Mini-Circuits ZX60-14012L amplifiers are used for monitoring the time-domain sub-THz (RF) response of the SIL setup. In the IF measurements, a Keysight 33512B arbitrary waveform generator (AWG) is programmed to deliver a nonlinear ramp signal to the AFM radar chip to generate a linearized sub-THz chirp signal. Also, the IF signal is detected using a Keysight DSOS 104A oscilloscope (IF Osc) followed by a Mini-Circuits ZFBT-6GW+ bias-T, an Analog Devices LTC6560 trans-impedance amplifier (TIA), and variable gain amplifiers (VGA) made of three cascaded Texas Instruments OPA2354 amplifiers. The differentiator block is implemented along the VGA cascade by adding a high-pass $R$-$C$ filter between stages. All the measurement instruments share a 10 MHz reference signal, and a trigger starts with the AWG ramp signal.

**Broadband chirp linearization.** Achieving the ultimate range resolution and range accuracy from the SIL AFM radar requires highly linearized chirp signals. Both short- and long-term errors in the system can cause the chirp nonlinearity. The short-term errors originate from the statistical phase variations in the system, like the VCO phase noise, which increase the IF noise floor and add skirt noise around frequency lines in the IF spectrum. However, the short-term errors are negligible for our applications for two reasons. First, the SIL process greatly decreases the VCO phase noise by the large stabilization factor of STC. Second, for short-range applications, which are our main focus, the transmitted and received signals have phase noise correlation and cancel each other during the down-conversion. The phase noise cancellation factor $C = 4sin^2(\pi f_\Delta \tau_d)$, derived in Ref. [41,45], expresses the amount of improvement obtained by this correlation, where $f_\Delta$ is the offset frequency away from the VCO oscillation frequency, and $\tau_d$ is the round-trip delay. For instance, for a target distance of 1 m ($\tau_d$ = 6.66 ns) and $f_\Delta$ = 1 MHz, the cancellation factor is 27.6 dB, demonstrating a substantial short-term error correction in the system.

On the other hand, long-term errors in the way of chirp linearization need special treatment to diminish their undesired effect on the IF spectrum. These errors broaden the width of IF frequency lines, especially for higher-order frequency lines. For this purpose, a two-step chirp linearization is adopted to remove the long-term systematic errors of the broadband chirp. The first step is proposed in Ref. [40], where a particle swarm optimization is deployed to linearize the chirp. This method mostly removes the errors raised by the nonlinear response of the VCO frequency versus the control voltage $V_{tune}$ by programming the AWG to generate a deformed ramp signal. This technique is very effective for lower-order frequency lines ($N \leq 2$) in the IF spectrum; however, it falls short of chirp linearization when we move to the higher-order frequency lines. The remaining small chirp nonlinearities are removed using the second step presented in Ref. [46], where a software-based technique is employed to resample the IF signal so that sampling is not performed at a fixed time interval but at a fixed phase interval. This step dramatically improves the chirp linearity for frequency lines N < 15. To achieve even better chirp linearization, we should devise new methods by combining hardware- and software-based techniques that are out of the scope of this article.

**Long-term frequency stabilization measurement.** To calculate the Allen deviation as a measure of long-term frequency stability of the radar's VCO (Fig. 2f), the instantaneous radiated frequency is measured over a long period as illustrated in Fig. S7 in Supplementary. For this measurement, instead of recording the signal frequency from the spectrum analyzer, we used the FFT of the down-converted RF signal at the output of the even-harmonic mixer displayed on the RF oscilloscope. The VCO frequency is measured over 20 μm time intervals (RBW: 50 kHz) for each point in Fig. S7. This method allows us to continuously measure the instantaneous frequency over short periods, which is much faster than the limited sweep time of the spectrum analyzer (> 4 ms).

**Range accuracy measurement.** The results reported in Extended Data Fig. 4a are obtained by the measurement setup in Extended Data Fig. 1a. For this measurement, we do not require the RF data; therefore, the horn antenna and the even harmonic mixer setup are removed, which enhances the SIL effect. A Thorlabs DDS300 direct drive translation stage with a capability of 0.1 μm minimum incremental movement is utilized for this measurement. To calculate the radar range accuracy, the corner reflector is moved with step size $dR$ = 1 μm for $R$ = 18.5, 25.4, and 45.4 cm, $dR$ = 2.5 μm for $R$ = 61 and 71 cm, and $dR$ = 5 μm for $R$ = 81 cm. The number of translation steps is 200 with 30 times IF signal recording at each position to find the error bars similar to Fig. 5a. Each error bar shows a deviation of ±2σ around the average value, equivalent to a 95.5% confidence factor. We should emphasize that, in this measurement, the phase of the first line in the IF frequency comb is used for range measurements[6].

**Data availability**

The data supporting the plots within this article and other study findings are available from the corresponding author upon reasonable request.

**Code availability**

High-level descriptions of the codes created to develop the mathematical models are available from the corresponding author upon reasonable request.

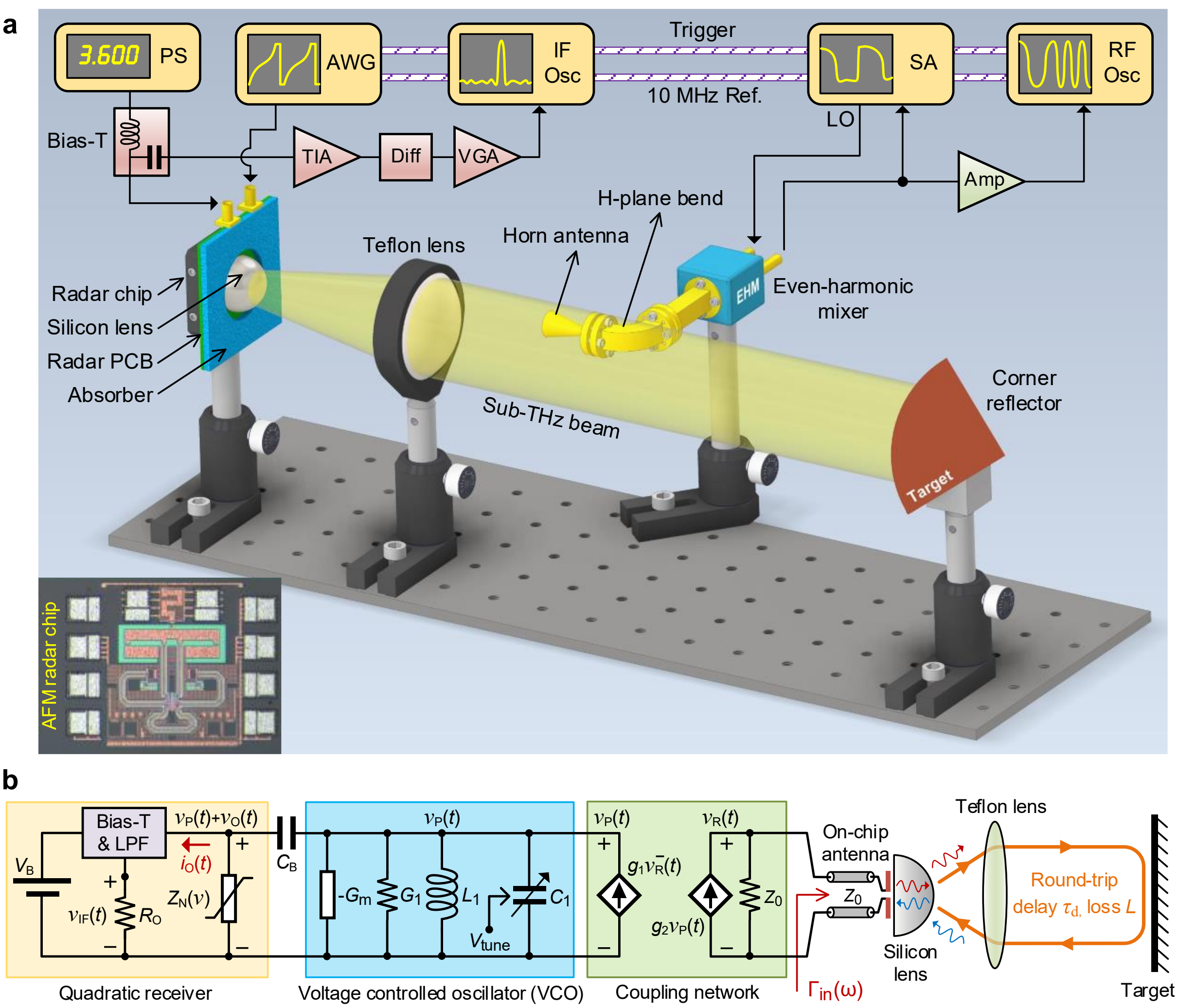

**Fig. 1 | Self-injection locking setup. a,** 3D view of the AFM radar measurement setup engaged in a SIL mechanism. The measurement instruments and RF/IF circuitry are illustrated for both time- and frequency-domain measurements. The fabricated AFM radar chip is shown in the inset. The VCO bandwidth at the second harmonic (220 GHz) is tunable between 54 GHz ($V_B$ = 3.6 V) and 68.6 GHz ($V_B$ = 2.9 V). The photograph of the measurement setup is displayed in the Extended Data Fig. 1a. **b,** The simplified schematic of the proposed AFM radar representing the self-injection locking. The actual circuit schematic of the radar is reported in Ref. [6]. PS: power supply. AWG: arbitrary waveform generator. Osc: oscilloscope. SA: spectrum analyzer. TIA: trans-impedance amplifier. Diff: differentiator. VGA: variable gain amplifier. Amp: amplifier. LPF: low-pass filter.

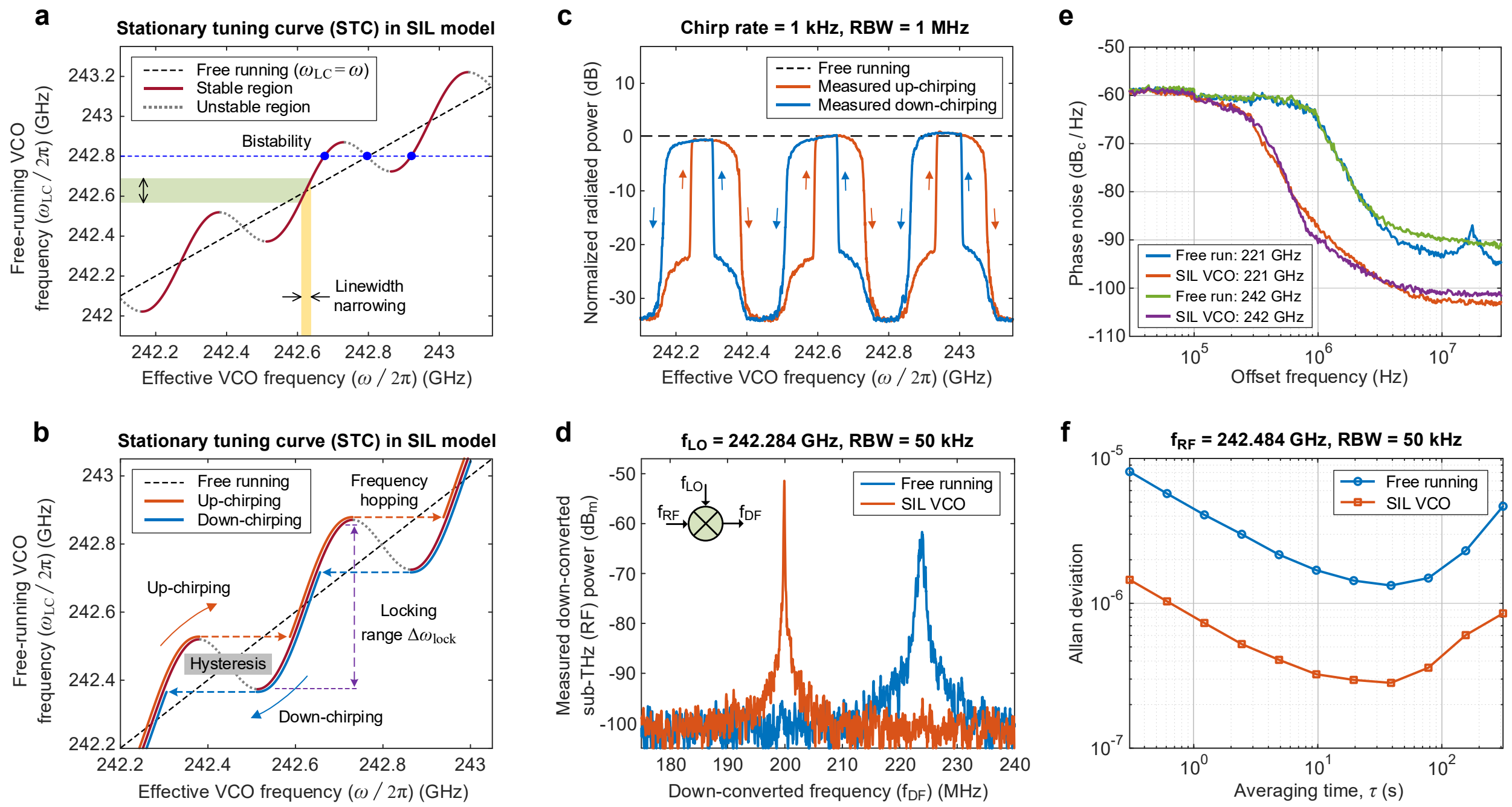

**Fig. 2 | Static analysis of SIL. a,** Stationary tuning curve (STC) of a VCO SIL to a free-space long delay line with length 42.8 cm ($\tau_d = 2.85$ ns). The control voltage ($V_{tune}$) defines the VCO frequency $\omega_{LC}$, and the sub-THz feedback from the free-space delay line determines the effective or actual oscillation frequency $\omega$. The effective frequency $\omega$ deviates from $\omega = \omega_{LC}$ (free-running case) when SIL happens. Within the locking range, the stabilization factor $d\omega_{LC}/d\omega > 1$, narrowing the oscillator linewidth. Bistability arises in parts of $\omega_{LC}$ where three solutions are available for $\omega$. Also, unstable areas of the STC are indicated by dotted lines. **b,** The zoomed STC with the demonstration of the instantaneous frequency transition between stable states at the turning points of the STC in up- and down-chirping. The spectral bistability of the SIL model illustrates a hysteresis behavior in up- and down-chirping. **c,** The normalized radiated power spectrums at sub-THz frequencies are illustrated for up- and down-chirping. The up- and down-chirping curves follow different trends due to the hysteresis in STC. **d,** Down-converted sub-THz (RF) spectrum of the free-running and SIL VCO shows narrower linewidth and stronger power for the SIL case. The sub-THz signal is down-converted by an LO signal at 242.284 GHz. For a better comparison, the two down-converted spectrums are intentionally shifted. **e,** Phase noise diagram of the free running and SIL VCO at 221 GHz and 242 GHz. **f,** Allan deviation diagram for the free running and SIL VCO at 242.484 GHz over 1000 s frequency measurements. Extended Data Fig. 1b shows the measurement setup for figures d, e, and f. RBW: resolution bandwidth.

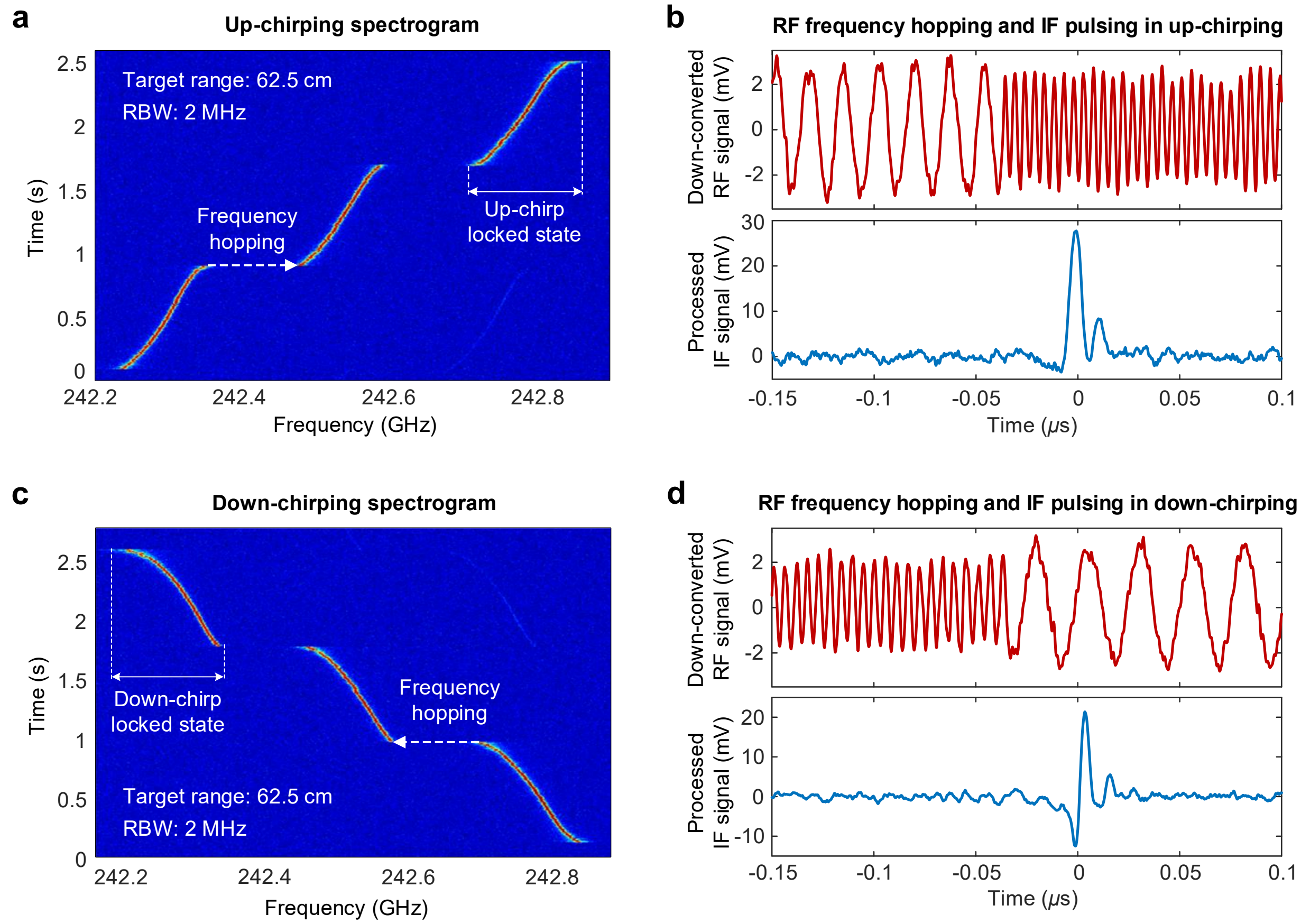

**Fig.3|Dynamic analysis of SIL. a,c,** Measured spectrogram of the sub-THz outputs of the SIL AFM radar for up- and down chirping, respectively. Here, the time axis is equivalent to $\omega_{LC}$ increment for up-chirping and $\omega_{LC}$ decrement for down-chirping. Frequency hoppings between adjacent stable branches and locked states are illustrated for both cases. **b,d,** Measured frequency hoppings of the sub-THz (RF) signals in the time domain are shown in red. Sub-THz signals are captured on an 8 GHz oscilloscope after down-conversion with LO at 242.284 GHz. The measured IF pulsings in the IF port of the SIL AFM radar are shown in blue. The time difference between RF frequency hopping and IF pulsing is related to the group delay of IF circuitry. The small trailing peak after the IF pulses can be attributed to strong frequency overshooting in the dynamic analysis (see Supplementary) or transient response of the IF circuitry. These IF pulses are achieved using a software-based method for pulse recovery, explained in the Supplementary. The down-chirping IF pulse magnitude is weaker and less stable than in the up-chirping case, with minor differences between the pulse shapes depending on the target distance.

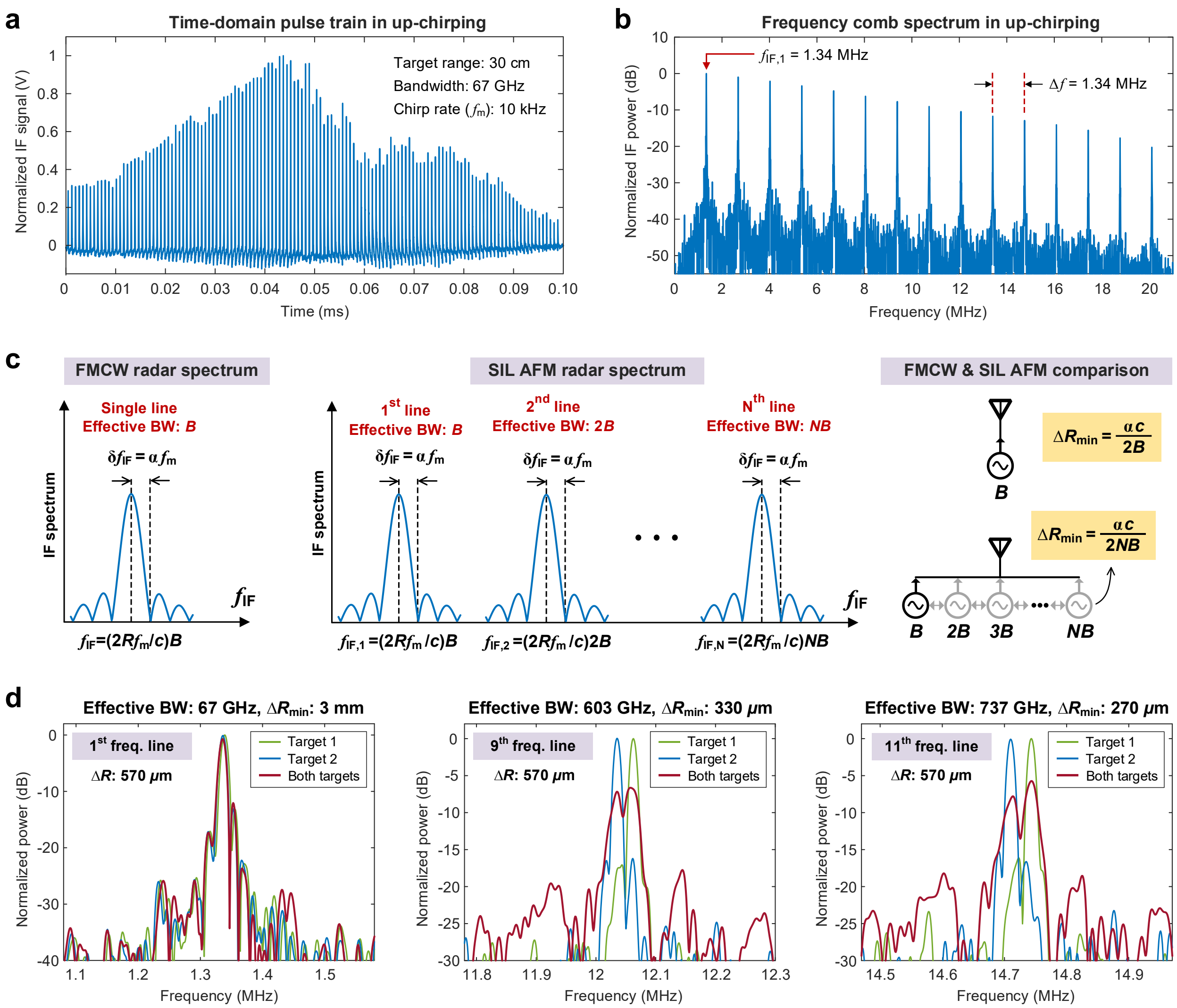

**Fig.4|Super-resolution ranging. a,** Time-domain IF pulse train in the up-chirping mode when SIL AFM radar sweeps a 67 GHz bandwidth ($B$) with a 10 kHz chirp rate. **b,** Frequency comb spectrum after taking FFT from the time-domain IF pulse train. The higher-order frequency lines in the IF spectrum are the harmonics of the first frequency line ($f_{IF,1}$ = 1.34 MHz), so their frequencies are $N$-times the first frequency line ($f_{IF,N} = N \times f_{IF,1}$). Here we have a target range: 30 cm, averaging: 100, chirp rate ($f_m$): 10 kHz, and window factor $\alpha = 1.33$. The power level drop at higher-order frequency lines is related to the frequency response of TIA and VGA at the IF port as well as the defects in the IF pulse recovery process explained in Supplementary. **c,** The comparison between an FMCW radar and the proposed SIL AFM radar. The SIL AFM operates as an ensemble of synchronized FMCW radars, where their effective bandwidths increase by the frequency line order ($NB$). **d,** Zoomed version of three frequency lines (1st, 9th, and 11th) in the IF frequency comb spectrum. Each graph includes three measurements when the SIL AFM radar looks at only the first target, only the second target, and both simultaneously. The targets are separated by $\Delta R$ = 570 μm at a 30 cm distance. Extended Data Fig. 2a shows the measurement setup for these results.

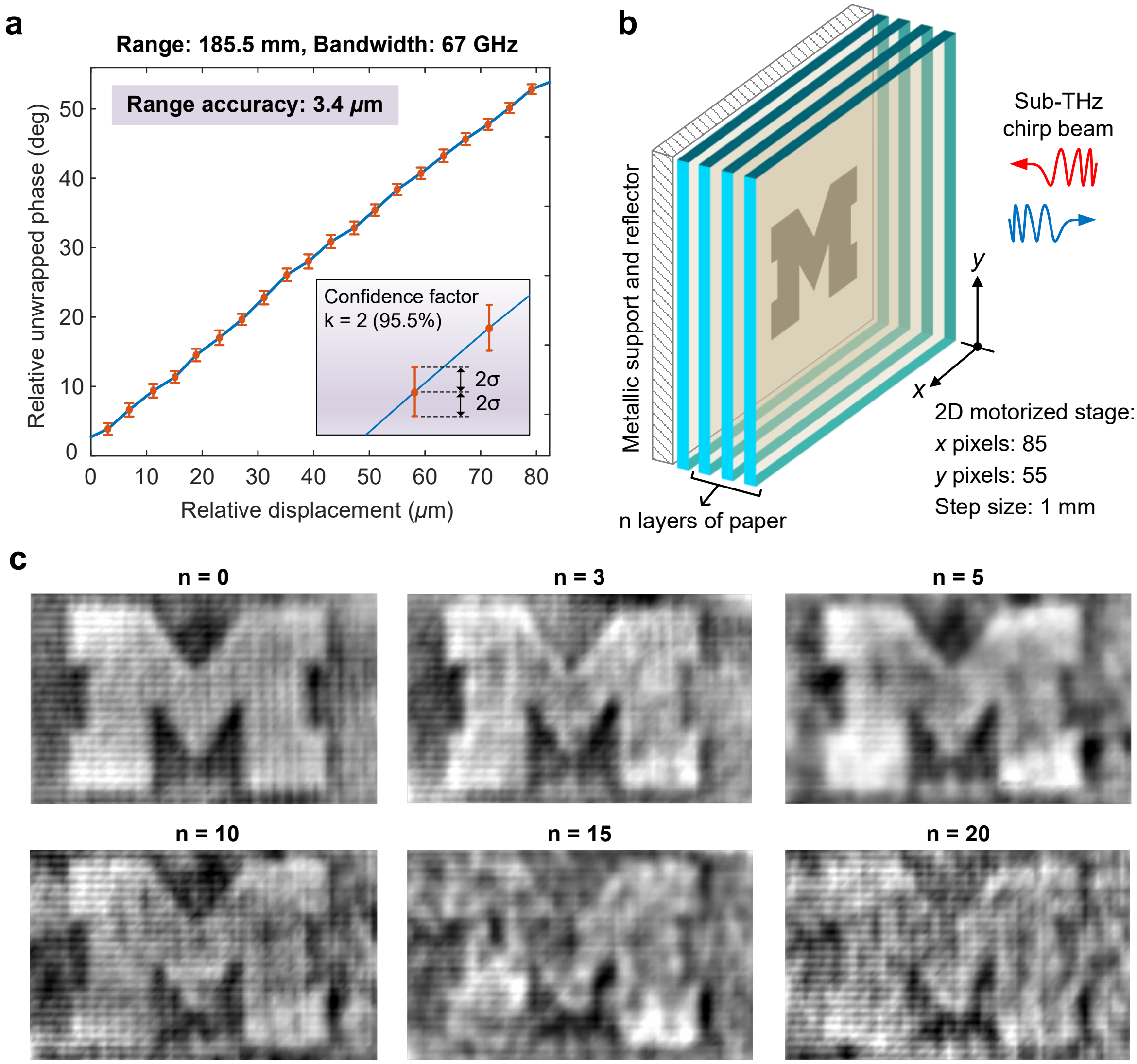

**Fig.5|High-accuracy ranging and high-precision imaging. a,** Range accuracy measurement results at 185.5 mm distance. A linear stage moves the targets with a 1 µm step, and we repeat 30 times the range measurement at each step. It shows SIL AFM radar has 3.4 µm range accuracy with confidence factor $k = 2$ (95.5%). In this graph, we report the range accuracy by measuring the phase of the first frequency line in the frequency comb spectrum. **b, c,** Imaging setup and image results from a printed "M" letter covered by $n$ layers of paper. The images are taken by the 2D moving of the image scene with 1 mm steps. The images are without applying image improvement methods. Extended data Fig. 2b displays the measurement setup of this imaging.

**a**

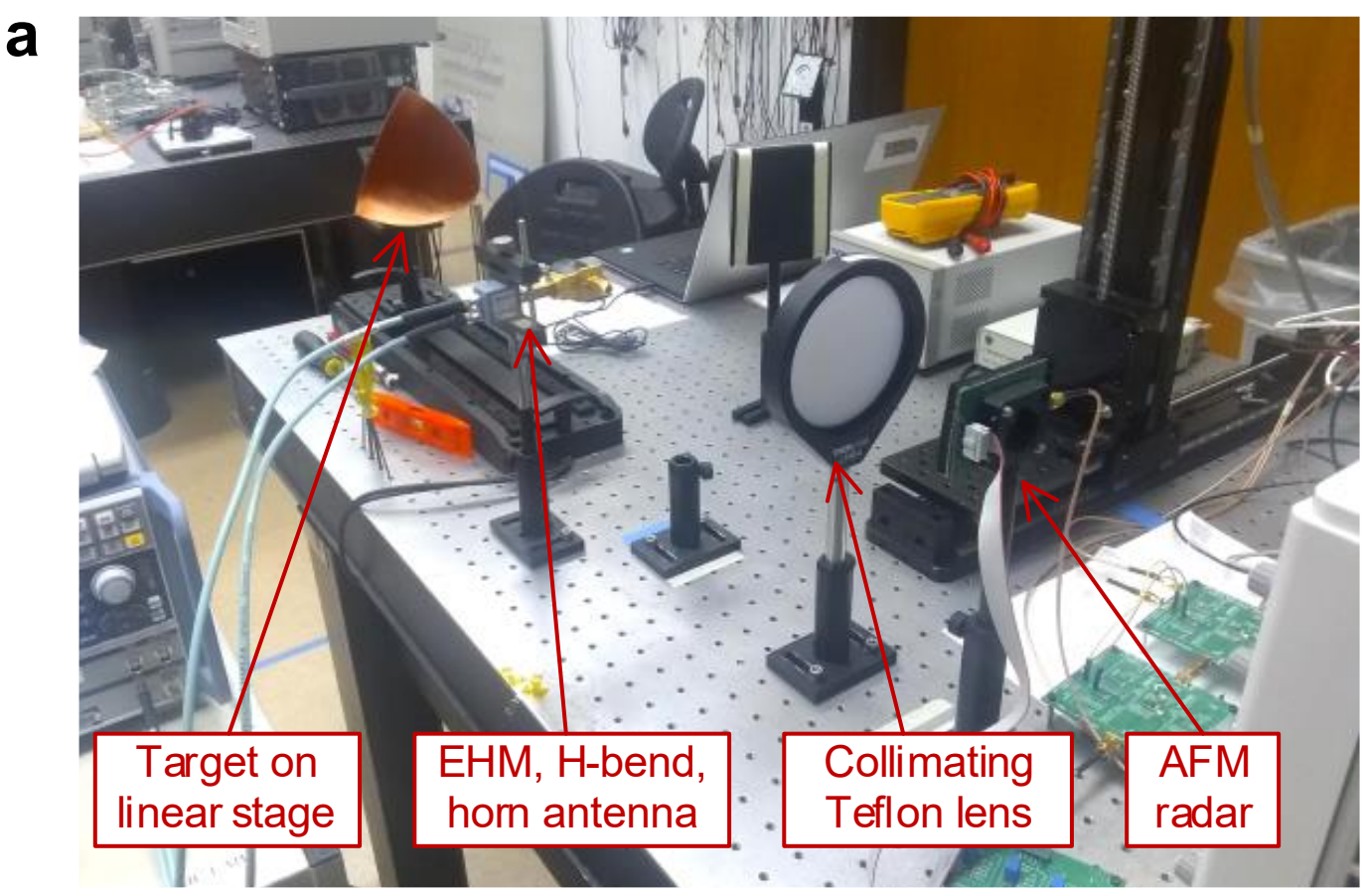

**b**

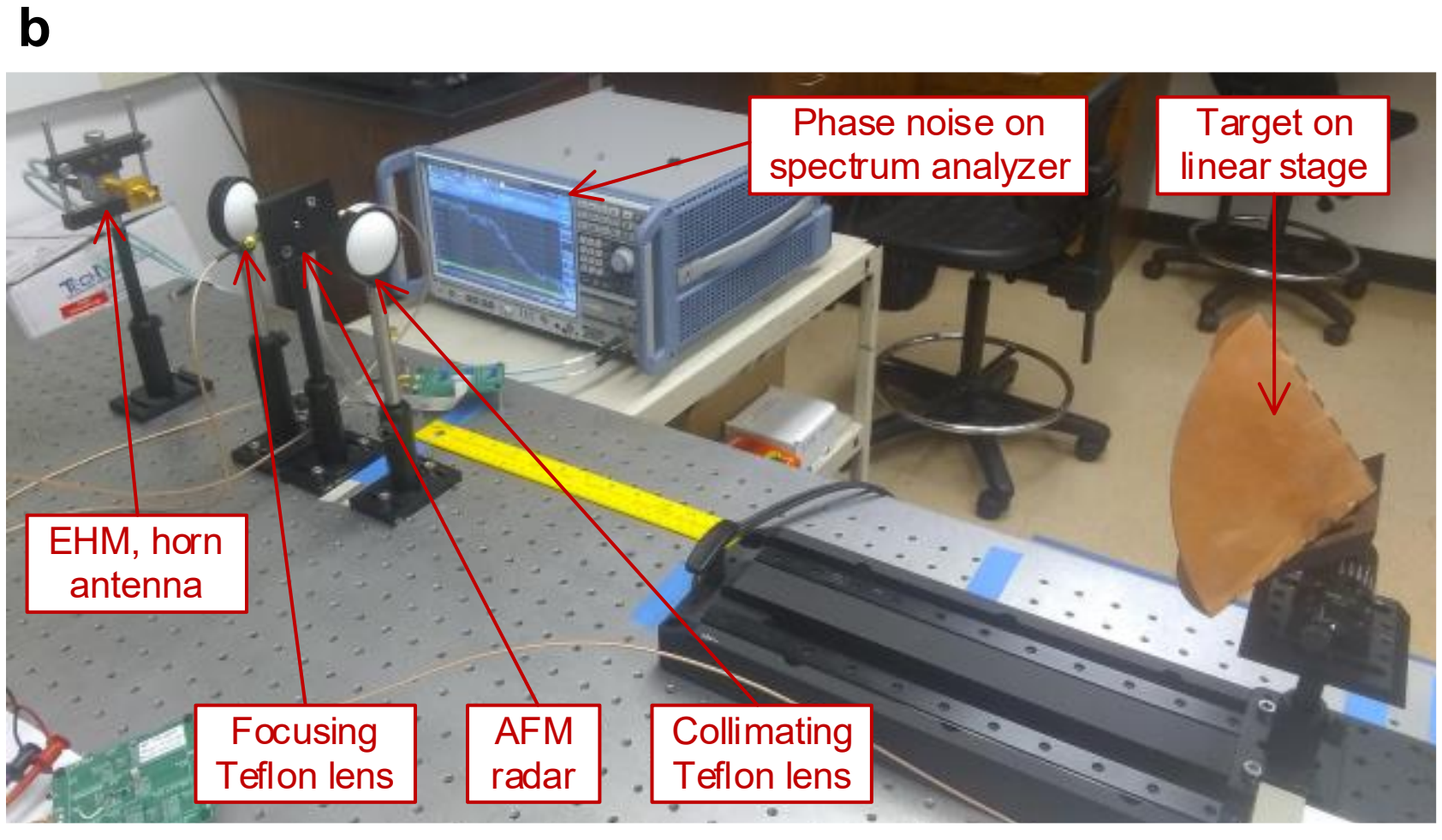

**Extended Data Fig. 1 | Photographs from SIL AFM radar measurement setups. a,** The practical implementation of Fig. 1a setup in the main text. This setup is also used for range accuracy measurement. The target is a corner reflector on a linear motorized stage. Even harmonic mixer (EHM), H-bend waveguide, and horn antenna take a sample from sub-THz radiated power of the AFM radar. **b,** Setup for single-tone sub-THz spectrum, phase noise, and frequency stability measurements (results reported in Fig. 2d, e, and f), where it utilizes a focusing lens to collimate the backside radiation of the radar chip into the aperture of the horn antenna for sub-THz measurements. The intrusion of the horn antenna in the way of target reflection in part (a) reduces the strength of the SIL; therefore, here, we use the backside radiation of the chip to remove this intrusion. However, due to the large power variations in the backside radiation measurement, it is only beneficial for single-tone measurements. For wideband measurements, we should employ the setup in part (a).

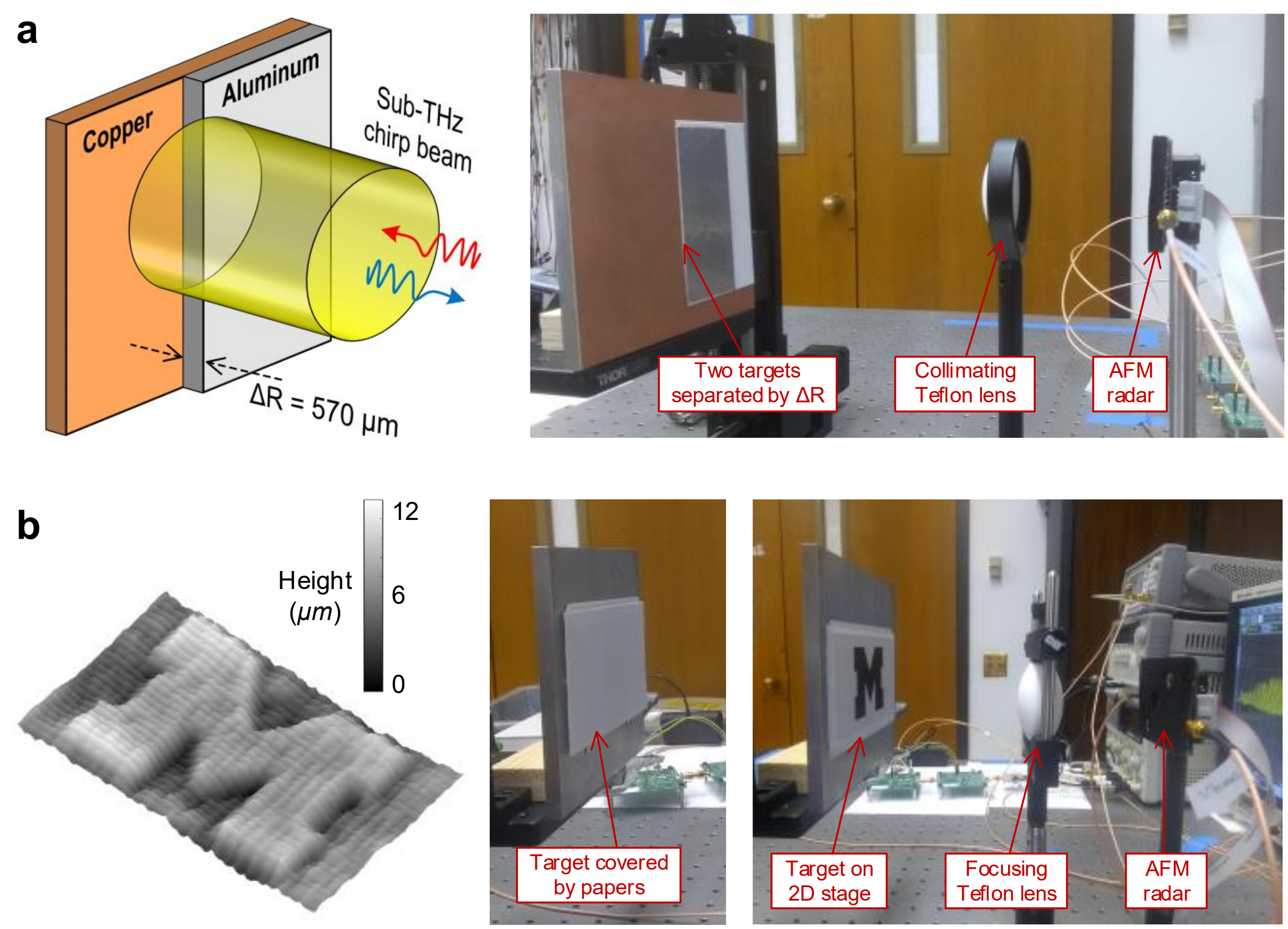

**Extended Data Fig.2|Photographs from range resolution and high-precision imaging setups. a,** Setup for range resolution measurement. An aluminum tape (70 μm thick) plus five layers of paper (each layer is 100 μm thick) forms a 570 μm step on a copper sheet support. The sub-THz collimated beam impinges on this step for range resolution measurement in Fig. 4d in the main text. **b,** Imaging setup from printed letter "M" with and without covering papers. A metallic support (target) behind the papers acts as an intensely reflective target. The 3D image of the printed letter "M" without paper covering ($n$ = 0) is reconstructed through a phase processing analysis of the first frequency line in the IF frequency comb spectrum.

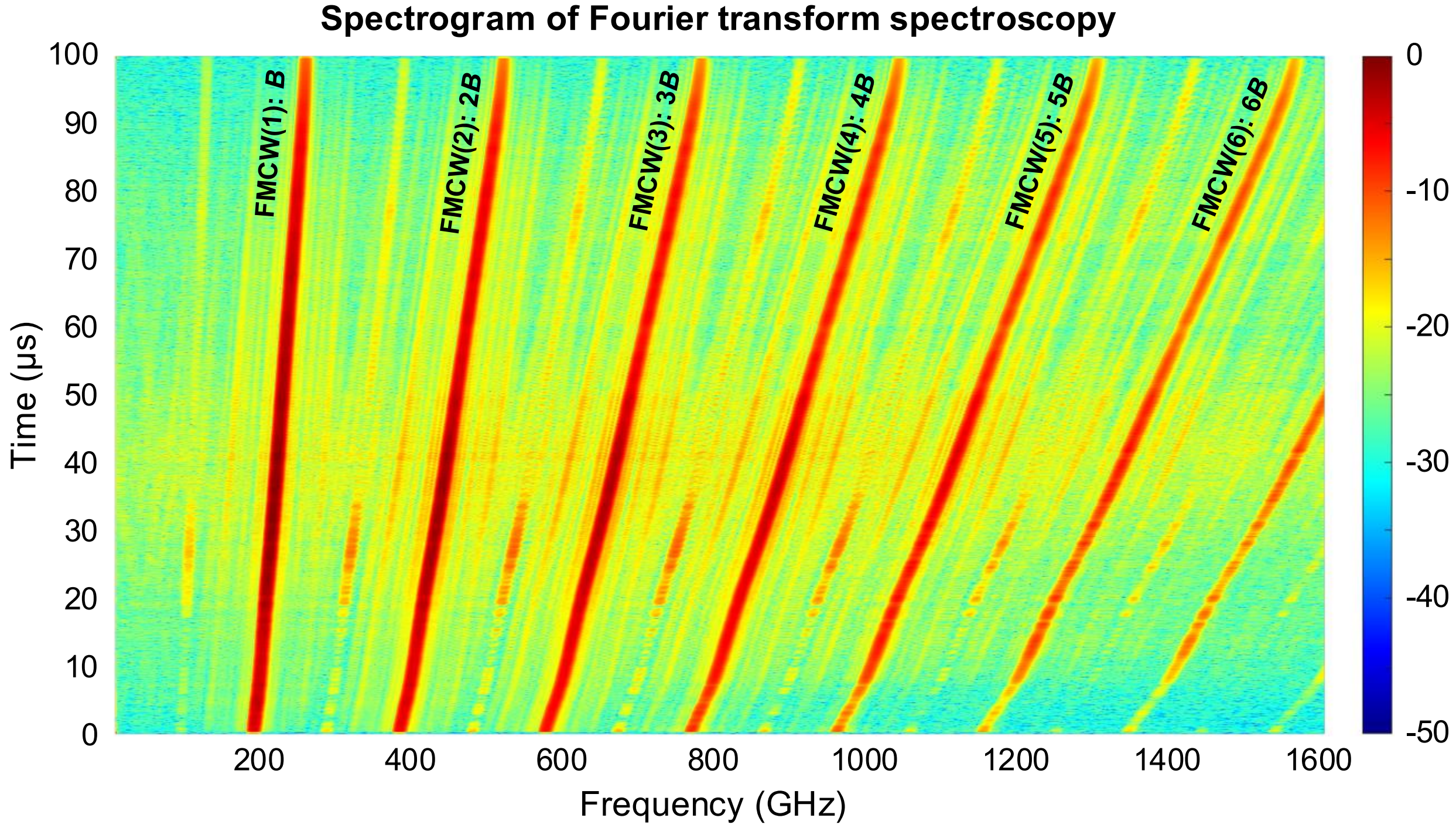

**Extended Data Fig.3|Fourier transform spectroscopy.** The normalized spectrogram of the SIL AFM radar, obtained using the FTS method, plots the radiated frequency during the up-chirping. The first radar FMCW(1) is the actual emitting frequency from 191-to-258 GHz ($B = 67$ GHz). The rest of the FMCW radars, FMCW($N \geq 2$) with expanded bandwidth $NB$, are fictitious and generated by the SIL nonlinearity.

**a**

**Range accuracy < 0.002%**

| Range (cm) | Range accuracy w/o averaging (μm) | Range accuracy w/ 100 averaging (μm) |
|---|---|---|
| 18.5 | 16.4 | 3.4 |
| 25.4 | 17.1 | 4.2 |
| 45.4 | 27 | 7 |
| 61 | 47.1* | 8.7 |
| 71 | 56.3* | 11.4 |
| 81 | 62.6* | 13.2 |

* These numbers are affected by the electrical noise of the translation stage in the IF frequencies.

**b**

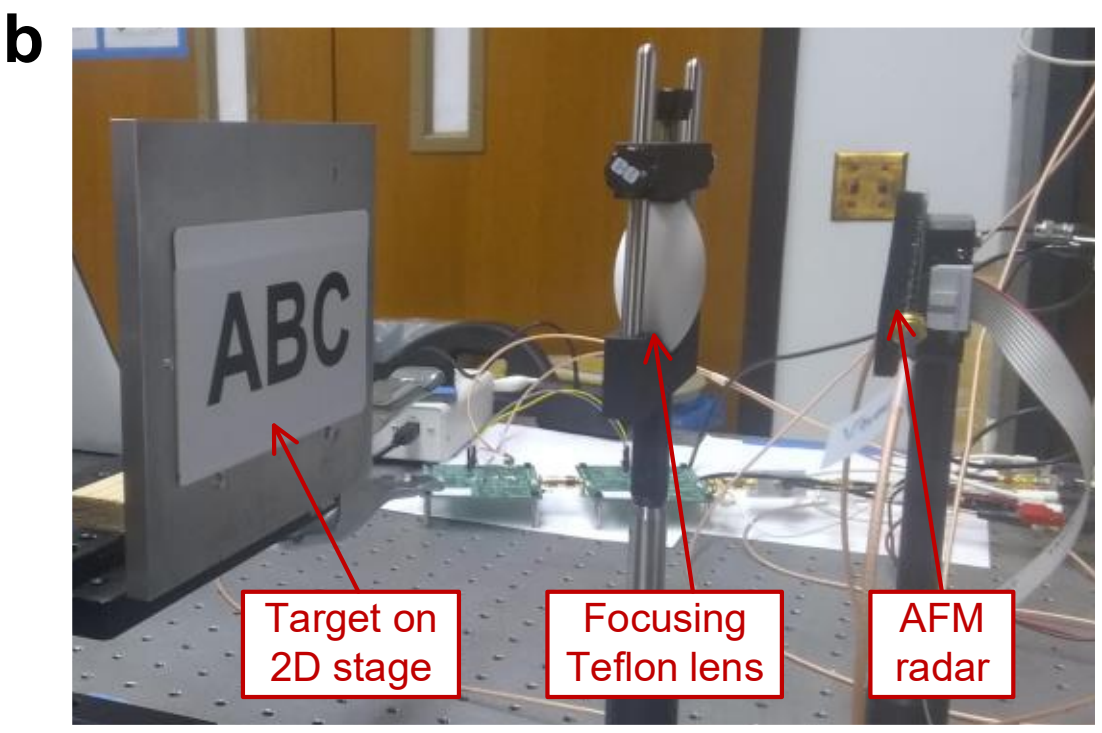

**c**

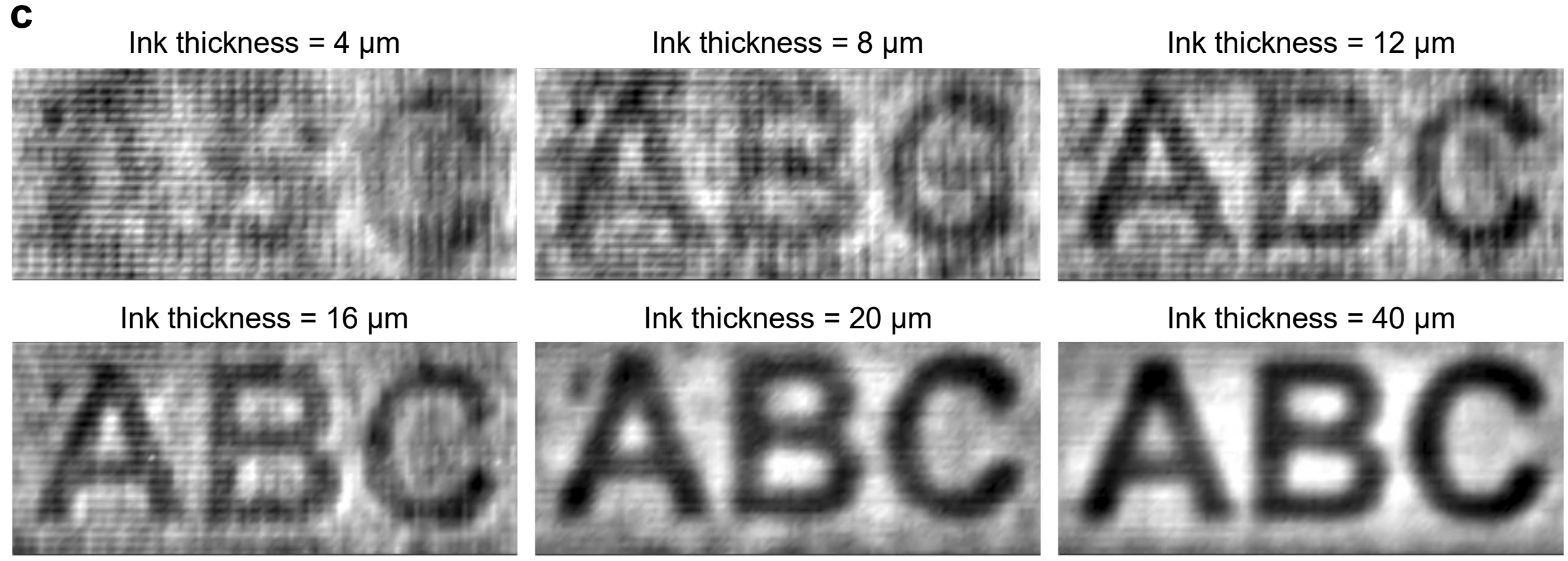

**Extended Data Fig.4|Range accuracy table with additional high-precision images. a,** Range accuracy table for several target ranges with and without signal averaging. Overall, the radar provides a range accuracy better than 0.002% with 100 times signal averaging. **b,** Imaging setup from the printed "ABC" letters with different ink thicknesses. **c,** Reconstructed images from "ABC" letters with different ink thicknesses with 100 times signal averaging. It shows the image quality improvement with thicker ink. For 4 μm ink thickness, it is hard to distinguish the "ABC" letters because 4 μm is close to the 3.4 μm range accuracy limit of the radar. Also, the micron-deep scratches on the metallic support involved in the range measuring deform the ink height measurement.

# Supplementary information:

# Super-resolution ranging using a sub-terahertz self-injection-locked frequency-modulated radar

S. M. Hossein Naghavi *et al.*

## I. Theory of self-injection locking (SIL)

The simplified circuit model in Fig. 1b in the main text describes the SIL AFM radar system through the two main signals: $v_P(t)$, the pump voltage on the voltage-controlled oscillator (VCO) representing the self-injection locking (SIL) mechanism, and $i_O(t)$, the output IF current of the quadratic receiver. The complete set of equations of motions are:

$$\frac{dV_{\mathrm{P}}(t)}{dt} = \left[j(\omega_{\mathrm{LC}} - \omega) - \frac{G_1}{2C_1} + \frac{G_{\mathrm{m}}\big(A_{\mathrm{P}}(t)\big)}{2C_1}\big(1 + j\alpha_{\mathrm{g}}\big)\right] V_{\mathrm{P}}(t) - \frac{g_1 g_2 Z_0 L}{4C_1} V_{\mathrm{P}}(t - \tau_{\mathrm{d}}) e^{-j\omega\tau_{\mathrm{d}}} \qquad \text{(S1a)}$$

$$i_{\mathrm{O}}(t) = G_m\big(A_{\mathrm{P}}(t)\big) b_g \frac{|V_P(t)|^2}{2} - C_1 b_c\, Re\left\{V_{\mathrm{P}}^*(t)\frac{dV_{\mathrm{P}}(t)}{dt}\right\} \qquad \text{(S1b)}$$

Equations (S1a) and (S1b) are obtained by applying KVL/KCL relations and slowly varying envelope approximation (SVEA)[1]. Here, $v_P(t)$ and $i_O(t)$ are real-valued functions of time, and $V_P(t)$ is a slowly varying complex-valued voltage amplitude, where $v_{\mathrm{P}}(t) = \frac{1}{2}V_{\mathrm{P}}(t)e^{j\omega t} + c.c.$ and $V_{\mathrm{P}}(t) = A_{\mathrm{P}}(t)e^{j\phi_{\mathrm{P}}(t)}$. In this form, $A_P(t)$ and $\phi_P(t)$ are slowly varying real-valued functions for the amplitude and phase of $V_P(t)$, respectively. In (S1) derivation, we assumed $v_P(t) >> v_O(t)$, where $v_O(t)$ is the output IF voltage forms across the nonlinear impedance $Z_N(v)$, responsible for frequency down-conversion in the quadratic receiver.

The VCO in Fig. 1b comprises an $LC$ tank made of inductance $L_1$ and tunable capacitance $C_1(V_{tune})$, with the loss term represented in the conductance $G_1$. The VCO tank has a resonant frequency of $\omega_{\text{LC}} = \frac{1}{\sqrt{L_1 C_1(V_{\text{tune}})}}$, controlled by voltage $V_{\text{tune}}$. The activity sustaining the VCO oscillation is represented by the negative conductance $-G_{\text{m}}(A_{\text{P}}(t))$, a function of voltage amplitude $A_{\text{P}}(t)$. Parameter $\alpha_{\text{g}}$ is the phase-amplitude coupling factor in the VCO[2] and is inspired by the Henry factor in laser diode operation[3]. The quantity $\alpha_{\text{g}}$ is small for our VCO, based on a comparison between simulation and measurement results; however, for the sake of completeness, we have included this term in (S1a). Part of the pump voltage at the VCO is extracted through the coupling network by the complex transconductance $g_2$, then transmitted and received by the on-chip antenna, experiencing a round trip with a total delay of $\tau_{\text{d}}$ and loss of $L$. Here, $Z_0$ is the characteristic impedance of the transmission line connecting the coupling network to the on-chip antenna. We assume the antenna and transmission line are perfectly matched.

Furthermore, in Fig. 1b, $v_{\text{R}}(t)$ is the voltage on the antenna port and can be expanded as the forward and backward voltages on the transmission line as $v_{\text{R}}(t) = v_{\text{R}}^{+}(t) + v_{\text{R}}^{-}(t)$. We can prove that the backward voltage is proportional to the delayed version of the pump voltage as $v_{\text{R}}^{-}(t) = -\frac{g_2 Z_0 L}{2} v_P(t - \tau_{\text{d}})$. Part of the reflected voltage $v_{\text{R}}^{-}(t)$ is injected into the VCO tank by the complex transconductance $g_1$, which is defined by the last term in (S1a), known as the SIL term. Without the SIL term, (S1a) represents the behavior of a free-running oscillator with the frequency of $\omega_{\text{LC}}$ determined by the $LC$ tank resonant frequency. However, with the SIL term, the VCO frequency changes to the actual or effective frequency $\omega$, which differs from the free-running frequency $\omega_{\text{LC}}$ as the reflected voltage affects the oscillator dynamics. Notably, (S1a) is similar to the Lang-Kobayashi equation in injection-locked laser diodes[4,5], where we use pump voltage on the VCO tank instead of the electric field inside a laser diode Fabry-Perot cavity.

Equation (S1b) is obtained by analyzing the quadratic receiver in Fig. 1b. Here, the quadratic nonlinearities originated from impedance $Z_N(v)$ performing the down-conversion to generate a kink ($i_O$) in the biasing current of the VCO during any frequency hopping. The impedance $Z_N(v)$ retains the nonlinear terms of the two components in the VCO tank, $-G_m$ and $C_1$, where the linear terms are involved in the VCO tank and nonlinear terms are included in the quadratic receiver. For brevity, we only consider the dominant quadratic terms in $Z_N(v)$ in our modeling. Here, we represent the resistive nonlinearity as $G_N(v) = -G_m b_g v$ and the capacitive nonlinearity as $C_N(v) = C_1 b_c v$, where $b_g$ and $b_c$ are the respective quadratic nonlinear coefficients for resistive and capacitive terms. In this simplified modeling, the nonlinear coefficient $b_g$ is the net nonlinearity of all transistors' transconductances in the actual circuit, and the nonlinear coefficient $b_c$ is the net nonlinearity of all varactors in the actual circuit, which includes collector-base and base-emitter junctions of bipolar transistors and part of the tunable capacitor $C_1$ that is attributed to the voltage across the capacitor, not control voltage $V_{tune}$. As shown in Fig. 1b, the voltage $v(t)$ across impedance $Z_N(v)$ is the summation of high-frequency pump signal $v_P(t)$ and low-frequency IF voltage $v_O(t)$. The bypass capacitor $C_B$ is considered to DC isolate the VCO and the quadratic receiver. The DC bias voltage $V_B$ can change the bias point of the nonlinear impedance $Z_N(v)$ and consequently adjust the nonlinear coefficients of $b_g$ and $b_c$. Finally, the IF current $i_O(t)$ is extracted through a bias-T and low-pass filter (LPF). The first term of the IF current $i_O(t)$ in (S1b) is related to the resistive nonlinearity $G_N(v)$, and the second term is attributed to capacitive nonlinearity $C_N(v)$. Notably, the second term on the right side of (S1b) is proportional to the time derivation of the first term. Also, resistor $R_O$ is considered to model the loading effects of the low-frequency components in the actual circuit, the bias-T, and the following trans-impedance amplifier (TIA) shown in Fig. 1a in the main text. For our circuit simulation, we model the bias-T with a large inductor $L_{\mathrm{BT}}$ and large capacitor $C_{\mathrm{BT}}$.

### A. Static analysis of SIL equation

The first step in analyzing the SIL equation of motion (S1a) is to perform a static analysis by considering $\frac{d}{dt} \to 0$. This analysis gives the stationary tuning curve presented in Fig. 2a,b in the main text. In the steady state, we can show $V_{\mathrm{P}}(t), V_{\mathrm{P}}(t-\tau_{\mathrm{d}}) \to V_{\mathrm{P}}$, where $V_{\mathrm{P}}$ is a complex constant number. Then, by separating the real and imaginary parts of (S1a), we obtain equation set (S2), where $\Gamma_{\mathrm{in}}(\omega) = -Le^{-j\omega\tau_{\mathrm{d}}}$ represents the reflection coefficient seen from the antenna port, as shown in Fig. 1b in the main text, capturing the propagation delay $\tau_{\mathrm{d}}$ and the loss $L$.

$$
\begin{aligned}
&-\frac{G_1}{2C_1} + \frac{G_{\mathrm{m}}(A_P)}{2C_1} + \frac{Z_0}{4C_1} Re\{g_1 g_2 \Gamma_{\mathrm{in}}(\omega)\} = 0 \\
&(\omega_{\mathrm{LC}} - \omega) + \frac{G_{\mathrm{m}}(A_{\mathrm{P}})}{2C_1}\alpha_{\mathrm{g}} + \frac{Z_0}{4C_1} Im\{g_1 g_2 \Gamma_{\mathrm{in}}(\omega)\} = 0
\end{aligned}
\tag{S2}
$$

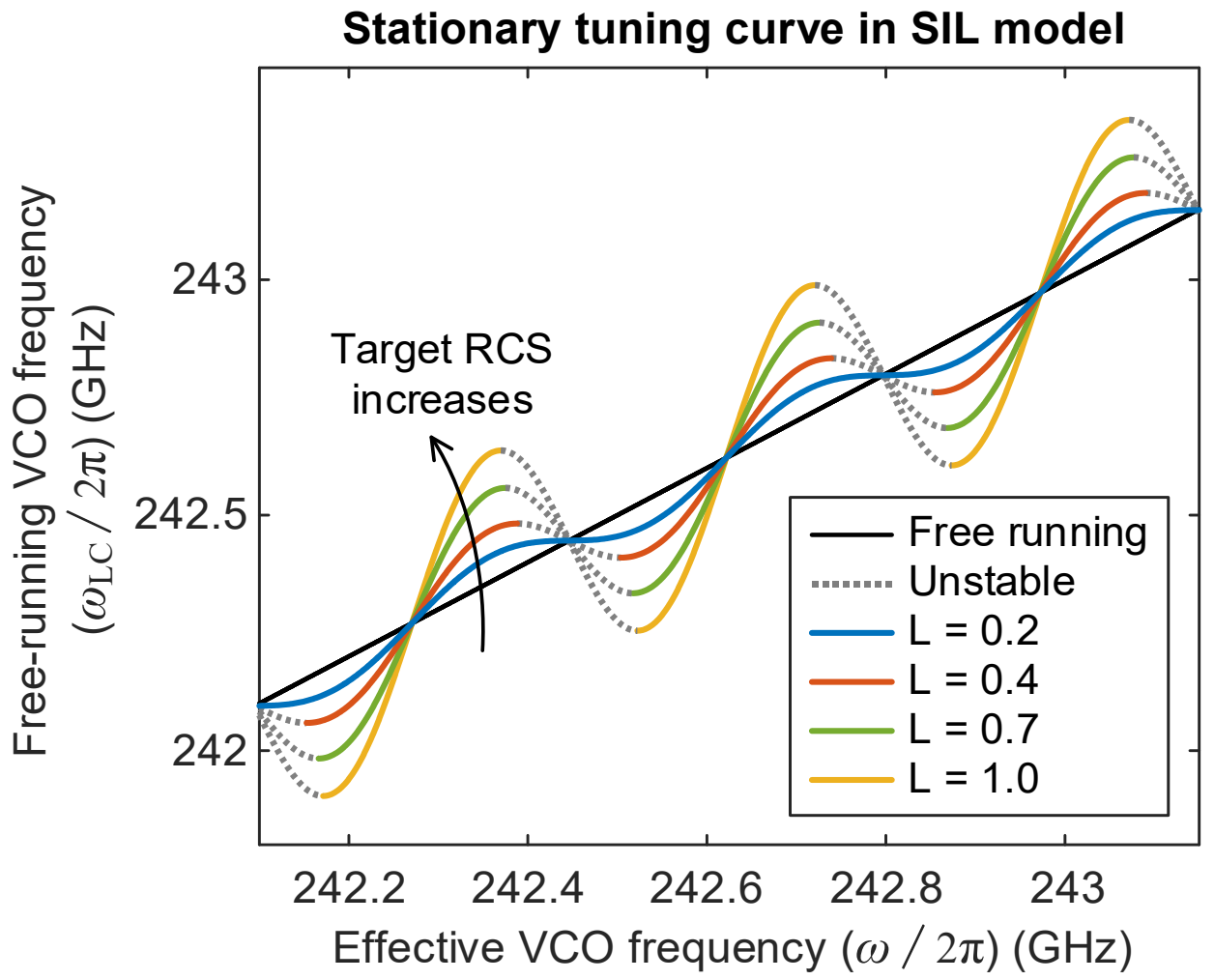

**Fig. S1|Static analysis of SIL system.** A simulated stationary tuning curve of VCO SIL for different propagation losses or equivalently target RCS is reported. Colored solid lines determine the stable areas; unstable regions are identified with gray dotted lines. Without a target when $L = 0$, the effective frequency is the same as free running frequency ($\omega = \omega_{\mathrm{LC}}$).

By eliminating $G_m(A_P)$ from (S2), we find (S3), the stationary tuning curve equation that demonstrates the relationship between the effective frequency $\omega$ and free-running frequency $\omega_{LC}$ of the VCO.

$$\omega_{\mathrm{LC}} = \omega - \alpha_{\mathrm{g}}\frac{G_1}{2C_1} - \frac{Z_0}{4C_1}\sqrt{1+\alpha_{\mathrm{g}}^2}|g_1 g_2 \Gamma_{\mathrm{in}}(\omega)|\sin\left(\psi - tan^{-1}(\alpha_{\mathrm{g}})\right) \tag{S3}$$

where $\psi = \measuredangle\{g_1 g_2 \Gamma_{\mathrm{in}}(\omega)\} = -(\phi_d + \omega\tau_{\mathrm{d}})$ and $\phi_d = -\measuredangle\{g_1 g_2\}$. It is noteworthy that the gain mechanism which arises from negative conductance $-G_m(A_P)$ is inessential to the stationary tuning curve. Fig. S1 illustrates the stationary tuning curve for different levels of loss term $L$, where $L$ is directly proportional to the target's radar cross section (RCS). By increasing RCS, VCO observes stronger SIL, which causes more significant variations in the stationary tuning curve. For a small value of $L \leq 0.2$, there is no bistable region on the tuning curve, which means no instability. However, for a larger $L > 0.2$, we observe bistable areas along the tuning curve that introduce unstable regions and are plotted by dotted lines. Solid lines in the figure demonstrate the stable branches of the tuning curve.

**Table S1 | Typical values for static analysis of VCO SIL used in reproducing Fig. S1 and Fig. 2a,b in the main text.** Capacitor $C_1$ is a function of $V_{tune}$, and its average value is reported here.

| Propagation loss | $L$ | 0 ~ 1 | VCO tank capacitance | $C_1$ | 43 fF |
|---|---|---|---|---|---|
| Propagation delay | $\tau_d$ | 2.85 ns | VCO tank conductance | $G_1$ | 6.6 mS |
| Magnitude of coupling terms | $\lvert g_1\rvert$, $\lvert g_2\rvert$ | 1.35 mS | Phase-amplitude coupling | $\alpha_g$ | -0.008 |
| Phase of coupling terms | $\phi_d$ | 1.7 rad | Antenna port impedance | $Z_0$ | 377 Ω |

Table S1 summarizes the typical values we used in plotting Fig. S1 and Fig. 2a,b in the main text. For this case, we assumed the VCO tank has $Q_{LC}$ = 10, and the target is 42.8 cm away from the radar. The antenna port impedance is intentionally chosen 377 Ω to be equal to the free-space impedance, then values of $|g_1|$, $|g_2|$, and $\phi_d$ are tuned to fit the measurement results in Fig. 2c in the main text.

### B. Dynamic analysis of SIL equation

The static analysis of (S1a) gives us a clear vision of the VCO SIL behavior in the stable branches of the tuning curve. However, to study the behavior of VCO in unstable areas of the tuning curve, we need to numerically solve the differential equation (S1a). Equation (S1a) is a delay differential equation (DDE) because of the $V_P(t-\tau_d)$ term on the right side. By Taylor expansion of $V_P(t-\tau_d) = \sum_{m=0}^{M} \frac{(-\tau_d)^m}{m!} \frac{d^m V_P(t)}{dt^m}$ to $M^{\text{th}}$ order, we can convert the DDE to an ordinary differential equation (ODE) with order $M$. For instance, Fig. S2 shows the simulation results of an up-chirp frequency hopping across an unstable region, considering $M = 2$ in the Taylor expansion. For this simulation, we use $\omega_{LC} = \omega_{LC1} + \beta t$ to scan the VCO free-running frequency, where $\omega_{LC1}/2\pi$ is the start frequency, and $\beta$ is the chirp rate. For the purpose of simulation, we first assume $\omega$ is a constant frequency $\omega_R$; then, we calculate the effective frequency from the temporal oscillation of slowly varying complex amplitude $V_P(t)$ using the $\omega = \omega_R + \frac{d\phi_P(t)}{dt}$ formula[2]. The value of $\omega_R$ is selected to fit the tuning curve of experimental results. Unlike the static analysis of VCO SIL, we need to define a gain mechanism to simulate the VCO behavior in the time domain. For this case, we can add a third equation to the equation set of (S1) to include the dynamic of the negative conductance $-G_m(A_P(t))$. However, for our analysis, we assume the time variation of negative conductance is much faster than the other processes in the equation set, so we use the steady state value of $G_m(A_P) = g_{mQ} \frac{2I_1(x)}{xI_0(x)}$, where $I_n(x)$ is the $n^{\text{th}}$-order modified Bessel function of the first kind, $x = \frac{A_P}{V_T}$, $V_T$ is the thermal voltage, and $g_{mQ}$ is the small signal transconductance of bipolar transistors in the VCO core[6].

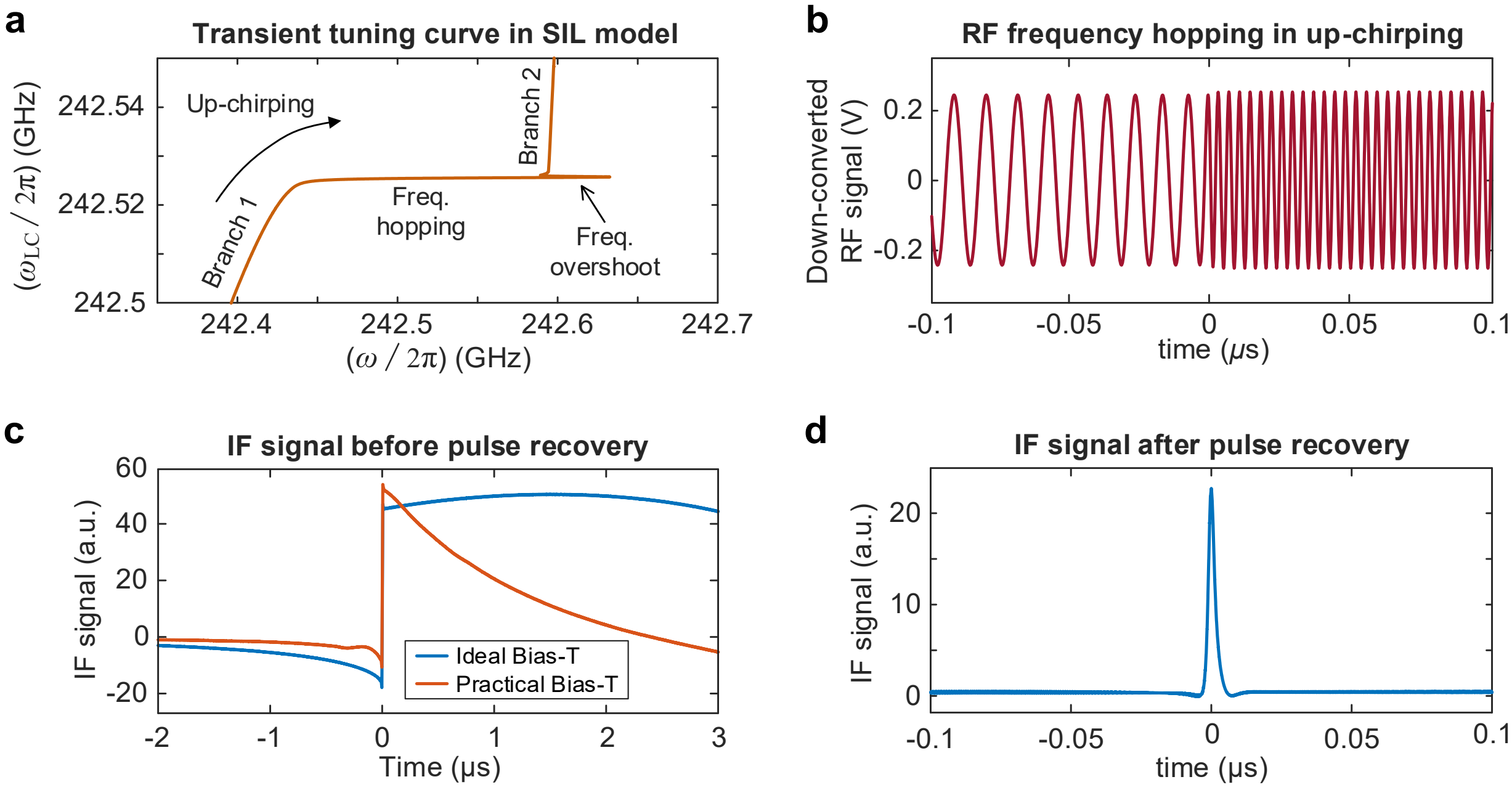

**Fig. S2|Dynamic analysis of SIL system. a,** Transient tuning curve in the SIL model shows the frequency hopping and overshooting between two adjacent stable branches during up-chirping. **b,** The sub-THz (RF) signal is down-converted by a 242.33 GHz LO signal and shows an immediate frequency variation during frequency hopping. **c,** Demonstration of IF voltage $v_{IF}(t)$ for the two cases with ideal and practical bias-Ts. **d,** Demonstration of IF signal $v_{IF}(t)$ after the pulse recovery process, which shows a sharp pulse synchronized with the RF frequency hopping.

Finally, Fig. S2a exhibits the transient tuning curve obtained from this calculation for up-chirping. It shows the instantaneous frequency hopping between two stable branches and the frequency overshooting before settling at Branch 2. Due to the 2nd-order Taylor expansion approximation, there are some differences between transient and stationary tuning curves, but this approximation still gives us an insight into the dynamic of VCO SIL at RF and IF frequencies when it jumps through an unstable region. Fig. S2b illustrates the immediate frequency change of the VCO signal $v_P(t)$ during the frequency hopping. For better visualization, the RF signal $v_P(t)$ is down-converted by a 242.33 GHz LO signal, similar to the results of Fig. 3b in the main text. On the other hand, at the IF frequency, the output signal $v_{IF}(t)$ is plotted in Fig. S2c for two cases with an ideal bias-T and a practical bias-T with parameters $L_{BT}$ and $C_{BT}$ reported in Table S2. For an ideal bias-T, the output IF voltage $v_{IF}(t)$ is close to a step function which relates to the kink current $i_O(t)$ in the biasing current of the VCO during frequency hopping[7]. However, for the practical bias-T, the output IF voltage shows a sharp change in the IF signal, followed by an exponential decay which is the step response of the bias-T. Fig. S6a compares the simulation and measurement results for the normalized IF voltages, which are in good agreement. Moreover, Fig. S2d depicts a sharp IF pulse after a recovery process explained in section II.C. This IF pulse recovery is essential in forming a frequency comb spectrum with broad bandwidth. Also, Fig. S2b,d illustrate that RF frequency hopping and IF pulse generation occur simultaneously.

**Table S2|Typical values for dynamic analysis of the SIL system represented in Fig. S2 and Fig. 3a,b in the main text.**

| Propagation loss | $L$ | 0.55 | Thermal voltage | $V_T$ | 26 mV |
|---|---|---|---|---|---|
| Chirp rate | $\beta$ | $7.54\times10^{14}$ rad/s$^2$ | Small-signal transconductance | $g_{mQ}$ | $5G_l$ |
| Effective frequency ($\omega/2\pi$) | $\omega_R/2\pi$ | 242.58 GHz | IF output resistor | $R_O$ | 0.8 Ω |
| Resistive nonlinear coefficient | $b_g$ | 0.01 ~ 1 | Bias-T inductor | $L_{BT}$ | 770 nH |
| Capacitive nonlinear coefficient | $b_c$ | 0 | Bis-T capacitor | $C_{BT}$ | 10.2 µF |

Table S2 shows the typical values we utilized in the dynamic analysis of the SIL system. For this case, we assume $L = 0.55$ to reproduce the stationary tuning curve of Fig. 2a in the main text. By studying several values for the nonlinear coefficients $b_g$ and $b_c$ and comparing them with measurements, we realize the resistive nonlinearity $G_N(v)$ is dominant in the down-conversion process, and we can ignore the effect of capacitive nonlinearity ($b_c = 0$). Moreover, for a range of $b_g$ values, we notice the only difference between generated IF pulses is their amplitude, and the pulse shape is unaffected by this term. Simulation results presented in Fig. S2 are obtained for $b_g = 0.01$. In addition, we consider the small-signal transconductance $g_{mQ}$ of bipolar transistors in the VCO core much larger than the loss in the VCO tank $G_l$ to deliver a strong pump signal.

## II. Additional measurements

### A. Effect of loss on the shape of IF pulses

One of the important features of the IF pulses generated through frequency hopping in VCO SIL is the robustness of pulses in maintaining an enduring shape with variations in the system. To study the robustness of generated IF pulses, we have simulated and measured IF pulses by placing lossy objects in front of the radar. Fig. S3a shows the simulation results for several attenuation terms $L$. As we can see, by increasing the loss in the system, the pulse amplitude decreases; however, the shape of the pulse remains constant. In the measurements, we placed several laminar materials with different thicknesses and loss tangents in the wave propagation path and monitored the pulse shape (See Fig. S3b). Similar to simulation results, the pulse amplitude weakens with higher loss but still, the form of the pulse remains unchanged. The trailing peak after each pulse in the measurements is due to the frequency overshooting effect (See Fig. S2a) and the transient response of the following IF circuitry (TIA and VGA). However, in simulations, due to the 2$^{nd}$-order ODE approximation, we don't see these trailing peaks.

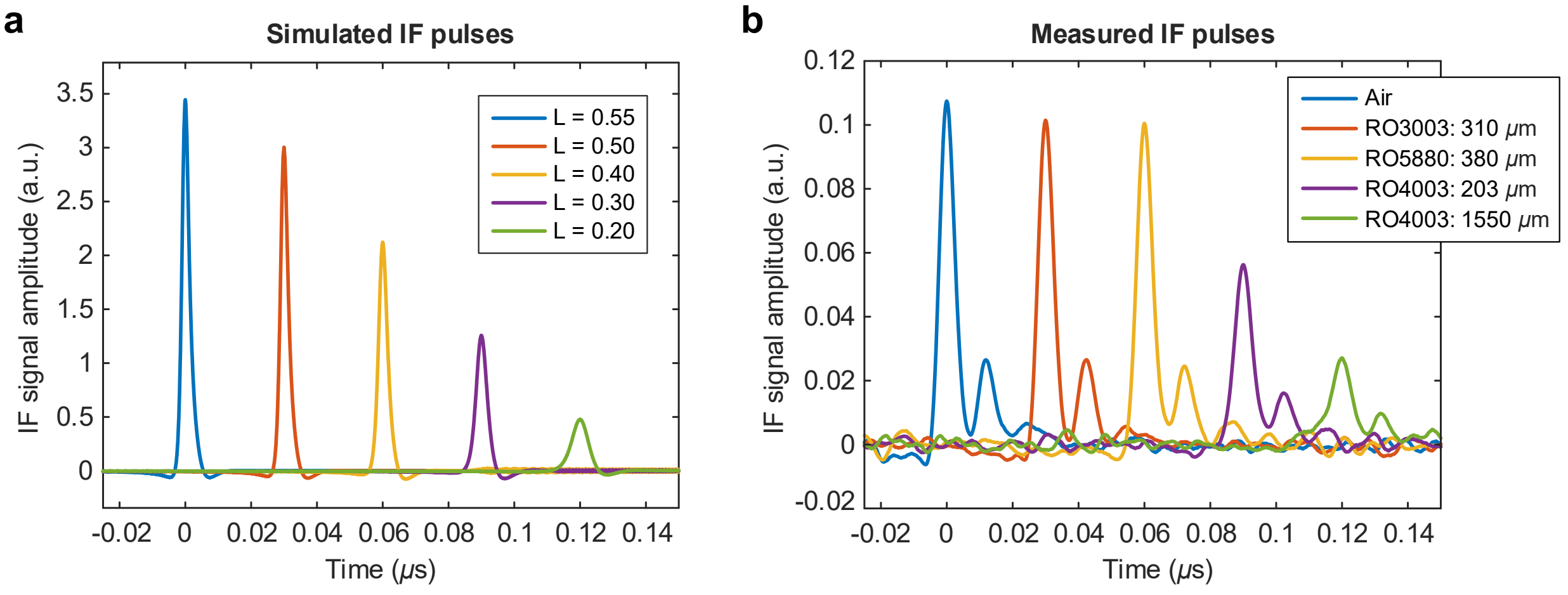

**Fig. S3|Robustness of IF pulse shape with respect to loss variations. a,** Simulated and time-shifted IF pulses for different loss term $L$ in the system. For this simulation, we assumed resistive nonlinear term $b_g = 0.01$. **b,** Measured and time-shifted IF pulses by adding lossy laminar plates in front of the radar with different thicknesses. The dielectric constant and loss tangent of these

materials are as follows: RO3003 (ε = 3, tanδ = 0.001), RO5880 (ε = 2.2, tanδ = 0.0009), RO4003 (ε = 3.55, tanδ = 0.0027). These IF pulses are recovered using a software-based technique discussed in section II.C.

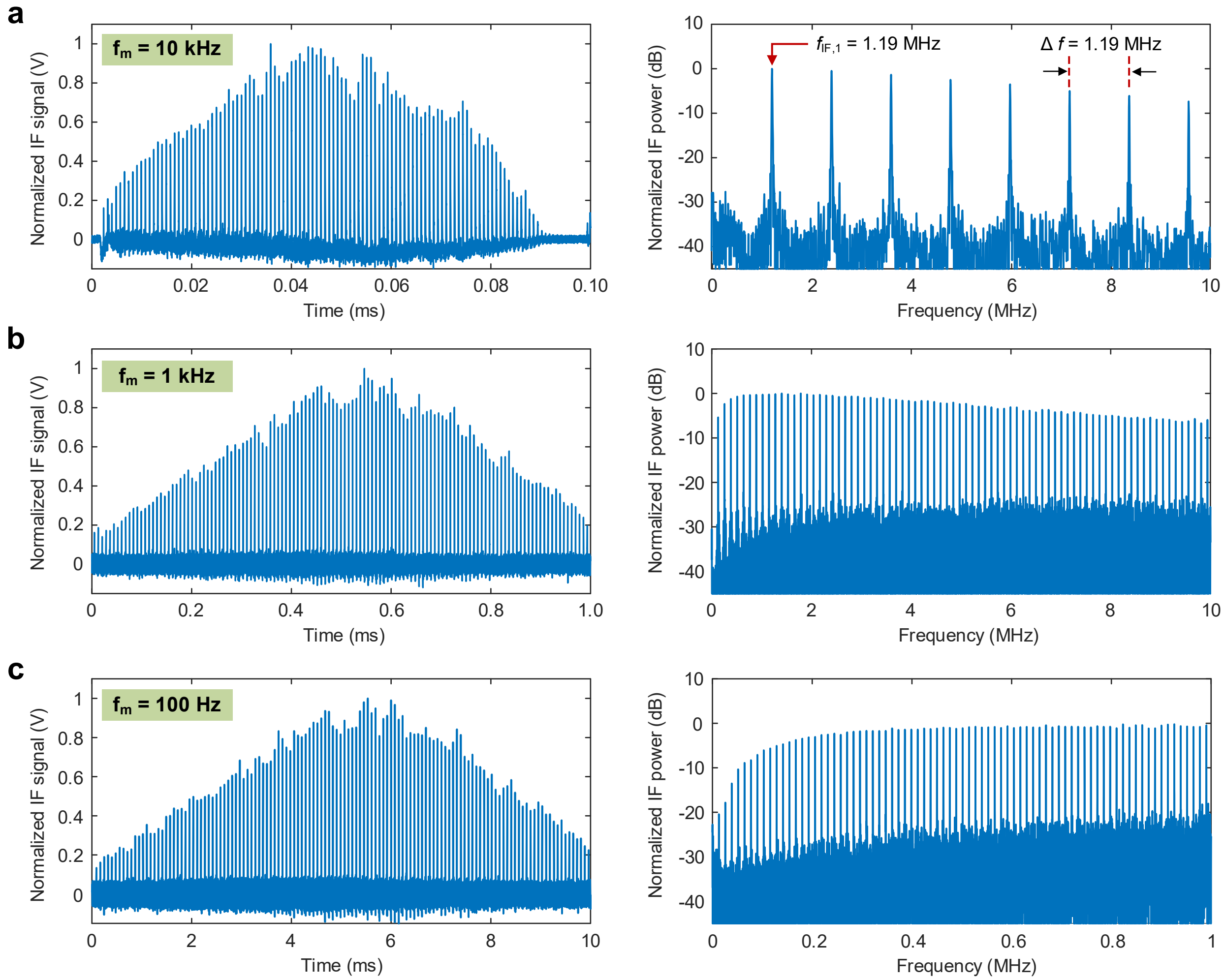

**Fig. S4|Improved range resolution by reducing the chirp rate.** The measurement results for time-domain IF pulse trains with associated frequency combs for a radar bandwidth of $B$ = 54 *GHz* and target distance of $R$ = 33 *cm*. **a,** Chirping frequency $f_m$ = 10 kHz. The failure in the pulse train at the two ends of the signal is related to the undesired behavior of TIA when the chirping rate is 10 kHz and above. **b,** Chirping rate $f_m$ = 1 kHz. The number of frequency lines is ten times larger than the case with $f_m$ = 10 kHz. The drop in amplitude of frequency lines at higher IF frequencies is related to the frequency response of TIA, VGA, and pulse recovery method explained in section II.C. **c,** Chirping rate $f_m$ = 100 Hz. The density of this frequency spectrum is 10 and 100 times more than the two cases of $f_m$ = 1 kHz and 10 kHz, respectively. We only plotted 1 MHz of the IF spectrum for a better demonstration of the separation between frequency lines. The amplitude drop of the first frequency lines is related to the frequency response of bias-T, which blocks frequencies below 100 kHz. The time-domain pulse train for all three cases has almost the same shape, and the only difference is the expansion of time separation between adjacent IF pulses. We should emphasize that the shape of IF pulses doesn't change by chirping frequency and always has a constant shape.

## B. Effect of chirp rate on the IF spectrum (frequency comb)

One of the exciting features of the SIL AFM radar is adjusting the required super-resolution by controlling the chirp rate $\beta = 2\pi B f_m$, where $B$ is the radar scanning bandwidth (or actual bandwidth), and $f_m$ is the chirping frequency. As mentioned in the main text, by moving to the higher frequency lines in the frequency comb spectrum, the achieved range resolution improves $N$-times, where $N$ is the frequency line order. Therefore, reducing the chirp rate can include more frequency lines in the IF spectrum and achieve better range resolution. However, the significant range resolution improvement comes with the cost of limited spacing between frequency lines, which puts constraints on the level of image scene complexity. Moreover, working with higher-order frequency lines requires sophisticated chirp linearization techniques due to their broader effective bandwidth. To show the adjustability of range resolution, we measure the time-domain IF pulse train and calculate the associated IF frequency comb spectrum for a fixed target at $R$=33 *cm* with three different chirping frequencies of $f_m$ = 10 kHz, 1 kHz, and 100 Hz, as illustrated in Fig. S4. Here, the time span is the reciprocal of $f_m$. As we can see from the figure, by reducing the chirping frequency $f_m$, we decrease the spacing between frequency lines in the frequency comb to enclose more frequency lines in a limited bandwidth of 10 MHz. For a better illustration of the frequency comb, in Fig. S4c, we only show the results for a 1 MHz window of the IF frequency comb because of dense frequency lines.

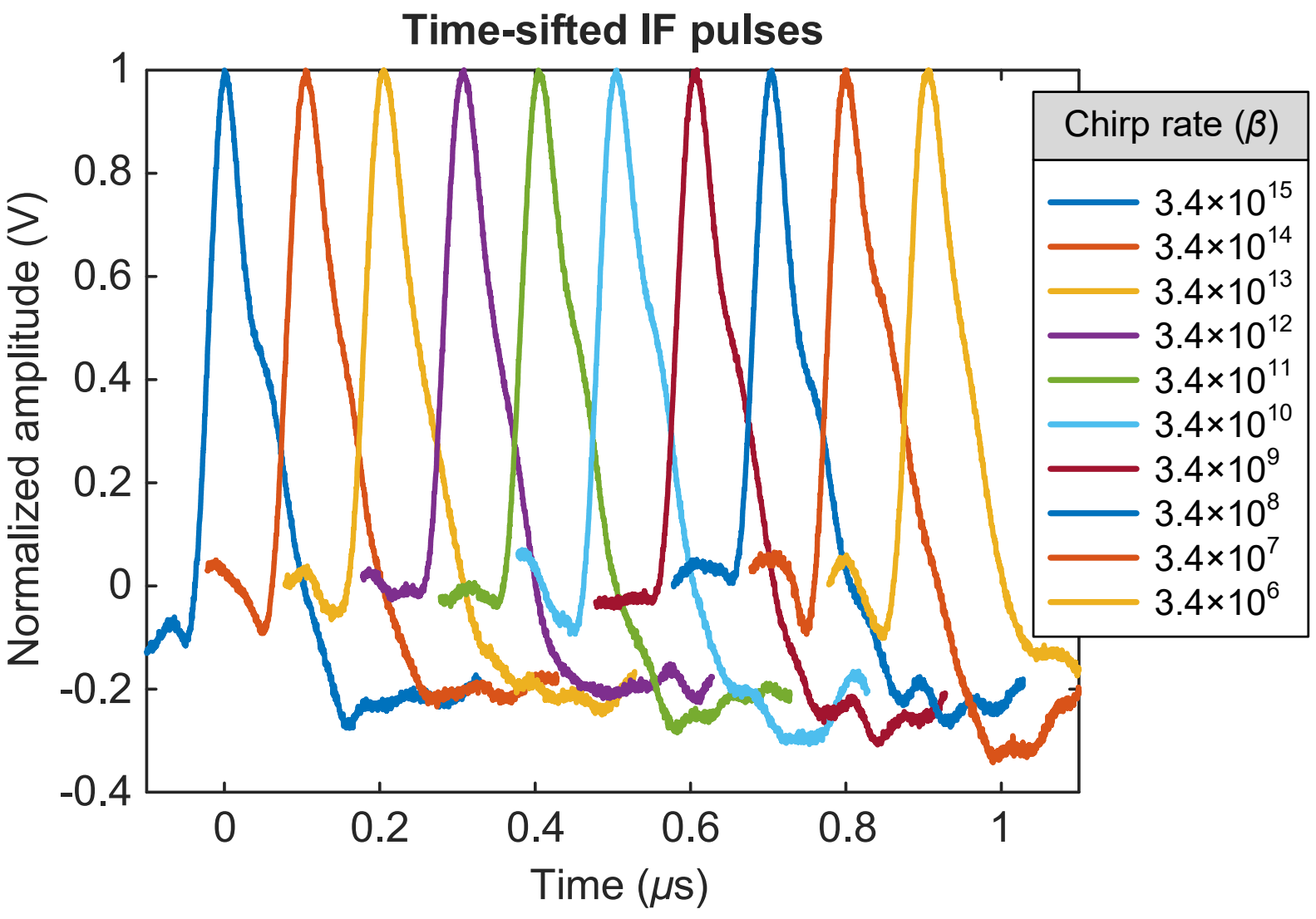

**Fig. S5|Constant pulse shape with respect to chirp rate.** The normalized measured IF pulses for several chirp rates are time-shifted for a better comparison, indicating the form of IF pulses remains unchanged. These IF pulses are recovered using a hardware-based technique discussed in section II.C.

We should emphasize that the adjustability of the range resolution in the SIL AFM radar by controlling the chirp rate is achieved due to the robustness of the frequency comb spectrum profile with respect to the variations of radar chirp rate. Accordingly, this property is related to the independence of the IF pulse shape to the variations in the system. To prove this point, measurements in Fig. S5 show that the IF pulse shape is independent of the chirp rate. Fig. S5 presents several measured IF pulses for different chirp rates and demonstrates the constant shape of pulses.

### C. IF pulse recovery process

As we predicted from our theoretical model in section 0 of Supplementary, the output IF signal $v_{IF}(t)$ is the step response of the bias-T, which shows a sharp transition in the IF voltage followed by an exponentially decaying tail. Fig. S6a compares the measured and simulated $v_{IF}(t)$ signals and demonstrates excellent agreement between measurement and simulation. However, to acquire the best performance from the super-resolution SIL AFM radar, we require a broadband frequency comb in the IF spectrum. Therefore, we need to remove the exponential decay of the IF signal and convert it to a sharp pulse. Fig. S6c shows the Bode diagram for the IF signal before pulse recovery, which rolls off by 20 dB/dec. This slope is equivalent to a dominant pole at very low frequencies and can be compensated by passing the IF data through a differentiator that adds a zero to the system's transfer function. The added zero crosses out the pole of the system. However, to prevent strong spikes caused by differentiation from noise, we pre-filter the data by a Gaussian window to reduce the variations raised by the noise. Fig. S6b shows the IF pulse recovery outcome that removes the exponentially tailing effect. Besides, Fig.S6c also exhibits the Bode diagram of the IF pulse after recovery, which has a flat spectrum, as expected from a sharp pulse.

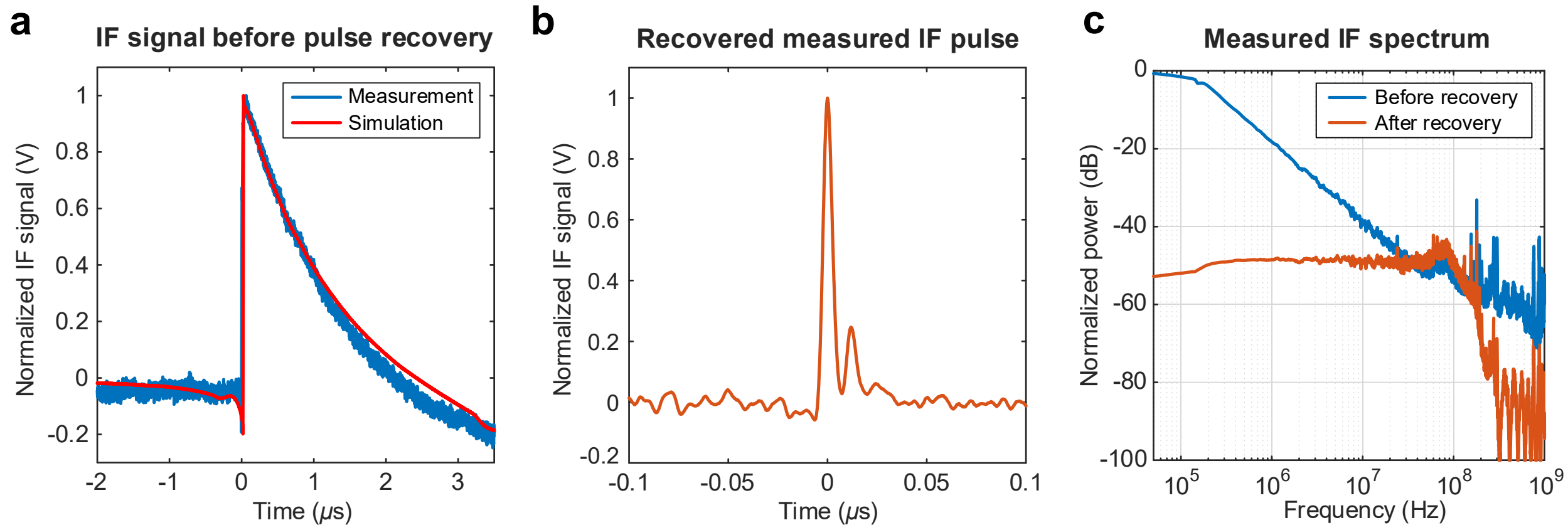

**Fig. S6|IF pulse recovery. a,** Comparison of measurement and simulation results for IF signals before pulse recovery. Simulation perfectly follows the measurement. **b,** Recovered measured IF pulse by smoothening the results of Fig. S6a with a Gaussian window, then taking a derivation. **c,** Bode diagram of measured IF pulses before and after recovery.

We should mention that the above algorithm for IF pulse recovery is a numerical method implemented with computational software. Nonetheless, we can accomplish the same task on the hardware level by adding a differentiator circuit after TIA, as depicted in Fig. 1a in the main text. The measurement results in Fig. 4 in the main text, Fig. S4, and Fig. S5 are obtained using a hardware-based technique. However, the implemented hardware differentiator is not as perfect as the software one; therefore, we observe a shallow drop in the amplitude of the IF frequency comb at higher frequencies in Fig. 4b and Fig. S4a,b. Also, the recovered IF pulses in Fig. S5 (hardware-based) are not as narrow as the one reported in Fig. S6b (software-based).

## D. Long-term frequency stabilization

In order to study the long-term frequency stability of the SIL VCO and compare it with the free-running case, the Allen deviation graph in Fig. 2f in the main text has been reported. The measured instantaneous frequency of this graph is shown in Fig. S7 for a 1000-second measurement. As we see, the frequency standard deviation for the SIL case is 5.3 times less than the free-running VCO.

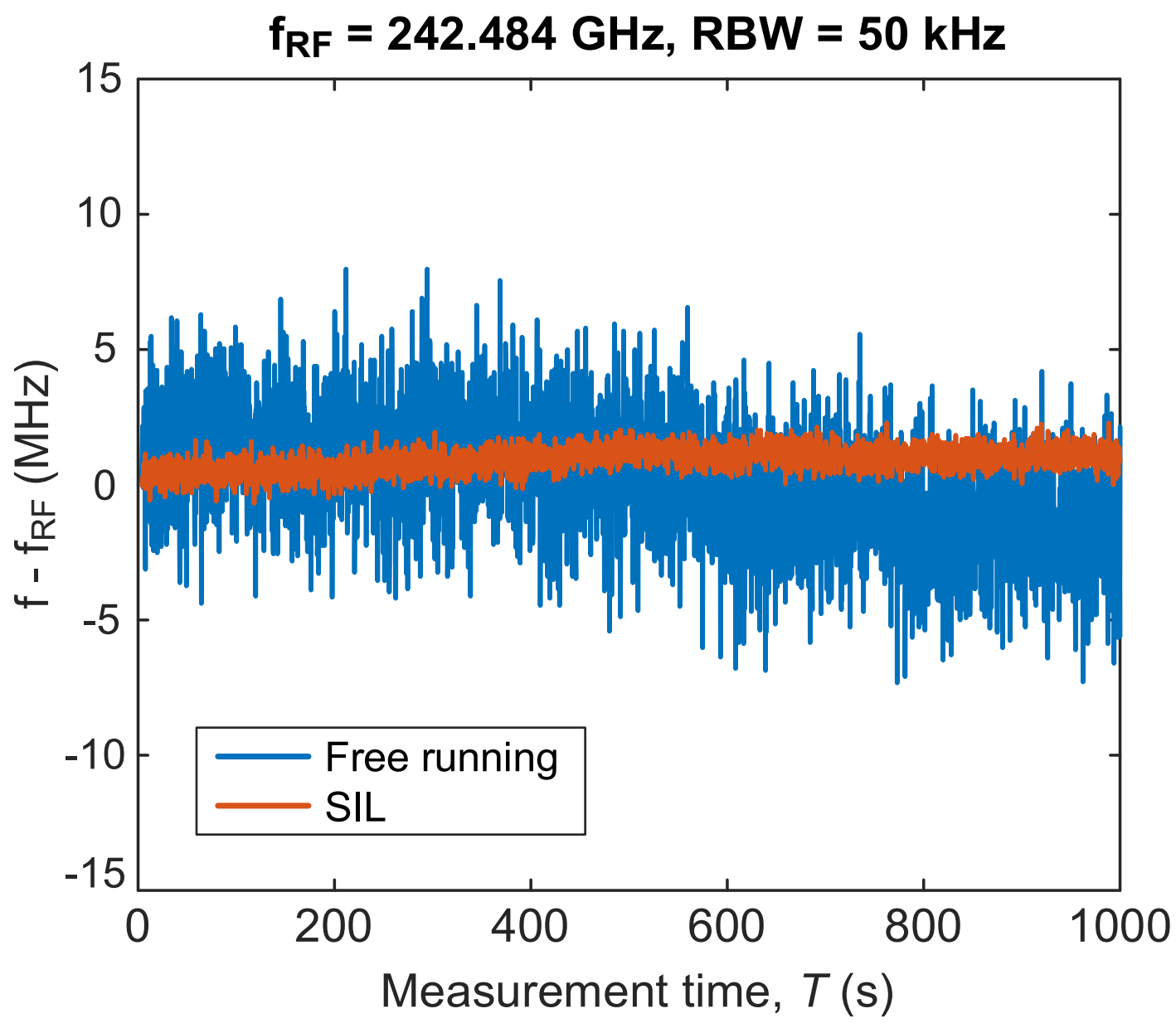

**Fig. S7|Frequency stabilization.** The instantaneous sub-THz (RF) frequency of the radar's VCO over a 1000-second measurement for both free-running and SIL cases. The resolution bandwidth for each frequency point is 50 kHz. The frequency standard deviation for free-running and SIL VCOs are 2.32 MHz and 0.44 MHz, respectively.

### E. Fourier transform spectroscopy (FTS) measurement setup

This section briefly explains the FTS method used in producing the Extended Data Fig. 3 in the main text. FTS is a well-known method for measuring the radiated frequency of coherent and incoherent sources by Fourier transformation of an interferogram produced by a two-beam Michelson interferometer[8]. In FTS, one of the interferometer beams remains fixed, and the length of the second beam is varied from zero to a maximum value. By constantly measuring the phase difference between the two beams with respect to the time delay difference between them, an interferogram is achieved, where its Fourier transform contains the radiated spectrum of the source. Notably, the operation of an FMCW radar is similar to the two-beam Michelson interferometry, as the fixed beam is the path inside the radar that takes a sample from the VCO and injects it to the LO port of the mixer, and the second path is the radiated signal to the air through TX/RX antennas. Therefore, constantly moving the target gives us a similar interferogram in FTS.

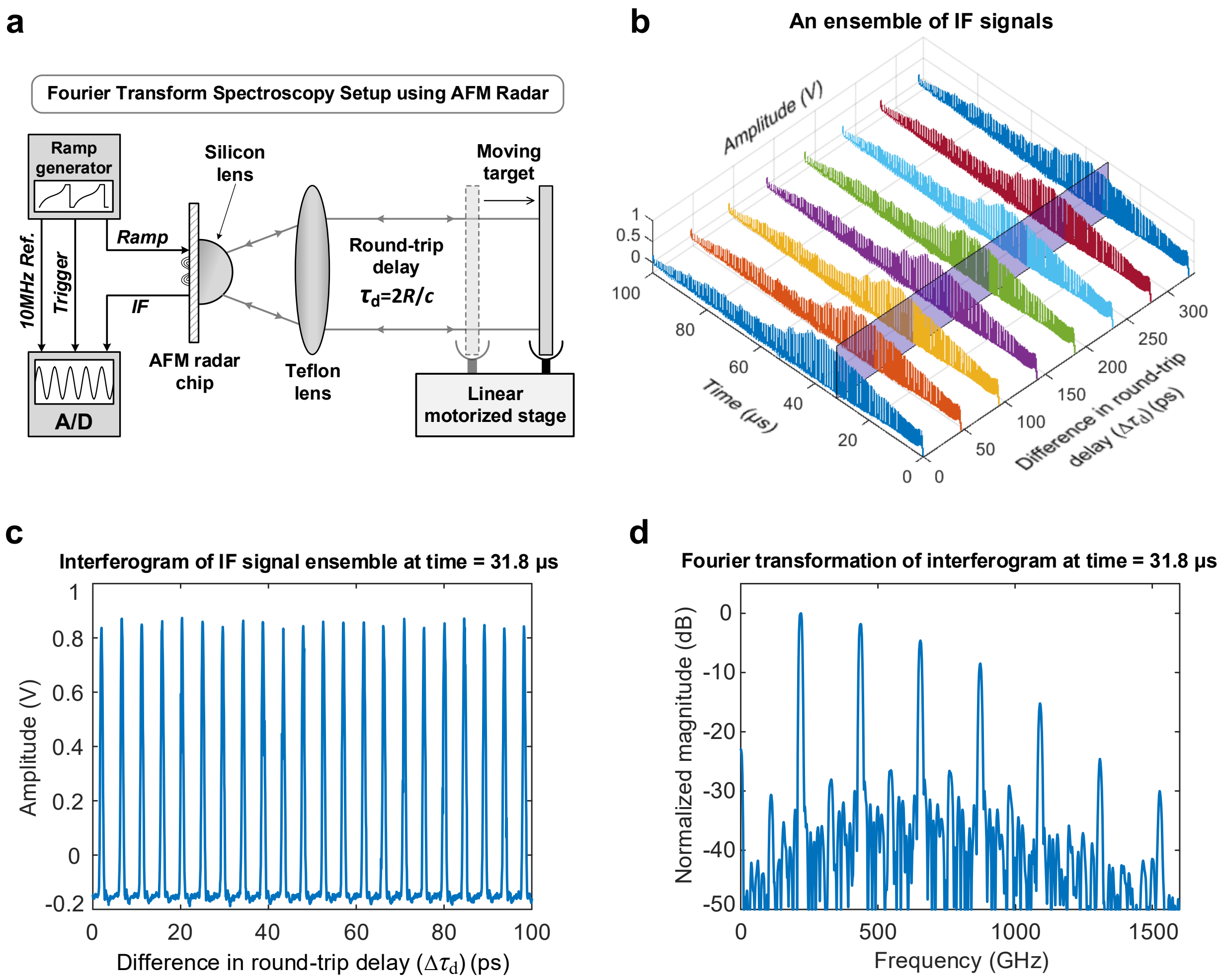

**Fig. S8 | Fourier transform spectroscopy measurement setup. a,** FTS setup using a SIL AFM radar, where the target is gradually moved by a linear motorized stage with 10 μm steps. The target is initially located at a distance of 40 cm and travels 5 cm away from the initial position. **b,** An ensemble of IF signals produced by putting the recorded IF signals for each stage position. Here, the difference between the round-trip delays ($\Delta\tau_d = 2\Delta R/c$) with respect to the initial target position is used instead of the absolute target position. **c,** The interferogram of the IF signal ensemble at a time of 31.8 μs, as depicted in part (b) by a cutting plane in the ensemble. **d,** A normalized Fourier transformation of the interferogram in part (c) shows a frequency comb deeply extended into THz frequencies.

Fig. S8a shows the FTS setup used for a SIL AFM radar where a linear motorized stage gradually displaces the target with 10 μm steps. At each step, we record the output IF signal, and by putting them side-by-side, we obtain an ensemble of IF signals, as illustrated in Fig. S8b. The IF signal ensemble is a 3D plot of IF amplitude versus time and the difference in round-trip delay $\Delta\tau_d$. Any cut from the IF signal ensemble at a fixed time demonstrates an interferogram containing the AFM radar's radiated spectrum at that specific time. For example, Fig. S8c shows one instance at time = 31.8 μs, a train of pulses separated by 4.5 ps. Therefore, the Fourier transform of this interferogram is a frequency comb with the first frequency line at 222.2 GHz, as shown in Fig. S8d. We should emphasize that due to the nonlinear mechanisms inside the SIL AFM radar, only the first frequency line is the actual radiated frequency from the chip, and the rest are fictitious radiations. Finally, by placing the Fourier transformations for all the times side-by-side, we acquire the spectrogram presented in the Extended Data Fig. 3. It obviously shows that the operation of SIL AFM radar is equivalent to the operation of an ensemble of synchronized FMCW radars, where the primary FMCW radar has a 67 GHz bandwidth from 191 GHz-to-258 GHz and the other fictitious FMCW radars emerge from the nonlinear processes inside the SIL AFM radar. The fictitious FMCW radars help achieve super-resolution, as discussed in the main text.